\documentclass[ALICE,manyauthors,nocleardouble]{cernphprep}

\usepackage{graphicx}
\usepackage{hyperref}   
\usepackage{flafter}
\usepackage{amsmath}
\usepackage{float}
\usepackage{wrapfig}
\usepackage{subfigure}
\usepackage{epsfig}
\usepackage{fancyhdr}
\usepackage{mathtools}
\usepackage{amssymb}
\usepackage{multicol}
\usepackage{multirow}
\usepackage{color}
\usepackage{lineno}
\usepackage{array}
\usepackage{soul}
\usepackage{caption}
\usepackage{enumitem}
\usepackage[titletoc]{appendix}
\usepackage{booktabs}
\usepackage[usenames,dvipsnames]{xcolor}

\usepackage{cite}

\newcommand{\kT}{\ensuremath{k_\mathrm{T}}}

\newcommand{\pT}{\ensuremath{p_\mathrm{T}}}

\newcommand{\pTleading}{\ensuremath{p_\mathrm{T}^\mathrm{track}}}

\newcommand{\rom}[1]{{\mathrm{#1}}}   

\newcommand{\kt}{{$k_{\rm T}$}}

\newcommand{\PbPb}{Pb--Pb}
\newcommand{\pp}{pp}
\newcommand{\sNN}{\ensuremath{\sqrt{s_\mathrm{NN}}}}
\newcommand{\GeV}{GeV}
\newcommand{\GeVc}{\text{\GeV/}\ensuremath{c}}
\newcommand{\MeV}{MeV}
\newcommand{\MeVc}{\text{\MeV/}\ensuremath{c}}
\newcommand{\pt}{\ensuremath{p_\rom{T}}}

\newcommand{\deltapt}{\ensuremath{\rom{\delta}\pt^{\rom{ch}}}}

\newcommand{\ptminmeas}{\ensuremath{\pt^\rom{min,meas}}}

\newcommand{\ptjet}{\ensuremath{p_\rom{T,jet}}}

\newcommand{\ptjetrawch}{\ensuremath{p_\rom{T,ch\;jet}^\rom{raw}}}

\newcommand{\ptjetch}{\ensuremath{p_\rom{T,ch\;jet}}}
\newcommand{\ptjetrec}{\ensuremath{p_\rom{T,\rom{ch\;jet}}^{\rom{det}}}}
\newcommand{\ptjetgen}{\ensuremath{p_\rom{T,\rom{ch\;jet}}^{\rom{part}}}}

\newcommand{\ptleadtrack}{\ensuremath{p_\rom{T}^{\rom{leading\;track}}}}
\newcommand{\pttrack}{\ensuremath{p_\rom{T}^{\rom{track}}}}

\newcommand{\chisq}{\ensuremath{\chi^{2}}}

\newcommand{\RCP}{\ensuremath{R_{\rom{CP}}}}

\newlength{\myfigwidth}
\newcommand{\TabH}[4]{
\begin{table}[h]
        \centering
                \begin{tabular}{#3}
                #4
                \end{tabular}
                \vspace{1ex}
                \caption{\small#2 \label{#1}}
\end{table}
}

\begin{document}

%
%
\begin{titlepage}
\PHnumber{2013-205}                 
\PHdate{02 November 2013}           
%
%
\title{Measurement of charged jet suppression \\ in Pb-Pb collisions at $\sNN=2.76$ TeV}
\ShortTitle{Measurement of charged jet suppression in Pb-Pb collisions at $\sNN=2.76$ TeV}   
%
\Collaboration{ALICE Collaboration%
         \thanks{See Appendix~\ref{app:collab} for the list of collaboration
                      members}}
\ShortAuthor{ALICE Collaboration}      
%

\begin{abstract}
A measurement of the transverse momentum spectra of jets in \PbPb{} collisions at
$\sNN=2.76$ TeV is reported. Jets are reconstructed from charged particles using
the anti-\kt{} jet algorithm with jet resolution parameters $R$ of
$0.2$ and $0.3$ in pseudo-rapidity $|\eta|<0.5$. The transverse
momentum \pt{} of charged particles is measured down to $0.15$ GeV/$c$ which
gives access to the low $p_{\rm T}$ fragments of the jet. Jets found
in heavy-ion collisions are corrected event-by-event for average
background density and on an inclusive basis (via unfolding) for
residual background fluctuations and detector effects. A strong suppression of jet
production in central events with respect to peripheral events is
observed. The suppression is found to be similar to the suppression of
charged hadrons, which suggests that substantial energy is radiated at
angles larger than the jet resolution parameter $R=0.3$ considered in the analysis. 
The fragmentation bias introduced by selecting jets with a high \pt{} leading particle,
which rejects jets with a soft fragmentation pattern, has a similar effect on the jet yield for central and peripheral events. 
The ratio of jet spectra with $R=0.2$ and $R=0.3$ is found to be similar in \PbPb{} and simulated PYTHIA \pp{} events, indicating no strong broadening of the radial jet structure in the reconstructed jets with $R<0.3$.

\end{abstract}
\end{titlepage}
\setcounter{page}{2}


\section{Introduction}
Discrete formulations of Quantum Chromodynamics (lattice QCD) predict
a phase transition to a new state of matter, the Quark-Gluon Plasma
(QGP), at an energy density above a critical value of about
1\,GeV/fm$^3$ and temperatures beyond $T_{\rm C} \approx
160$ MeV \cite{Karsch:2003jg,Bazavov:2011nk}.  In this state, the
elementary constituents of hadronic matter, quarks and gluons,
are deconfined and chiral symmetry is expected to be restored.
The conditions to create a QGP are expected to be reached for a short time (few fm/$c$) in the overlap
region of heavy nuclei colliding at high energy.

One of the tools to study the properties of the QGP is provided by
hard (large momentum transfer $Q^2$) scattering processes of the partonic constituents
of the colliding nucleons.  
These hard scatterings occur early in
the collision ($\ll 1\,\text{fm}/c$) and 
the outgoing partons propagate through the expanding hot and dense medium and fragment into jets
of hadrons. Jet fragmentation in heavy-ion collisions is expected to be modified (relative to the parton
fragmentation in the vacuum) due to parton-medium interactions, e.g.\ radiative and collisional parton
energy loss (jet quenching) \cite{Gyulassy:1990ye,Baier:1994bd}.
The initial hard parton production cross sections are calculable using
perturbative QCD (pQCD) and the non-perturbative vacuum fragmentation process can be well calibrated
via jet measurements in elementary collisions. 

Jet quenching has been observed at RHIC
\cite{Adcox:2001jp,Adler:2003qi,Ada03b,Adams:2003im,Arsene:2003yk,Back:2003qr} and at the LHC
\cite{Aamodt:2010jd,Aamodt:2011vg,ATLAS-CONF-2012-120,CMS:2012aa,Aad:2010bu,Chatrchyan:2012nia,Aad:2012vca} via the measurement of high-$\pt$ inclusive
hadron and jet production, di-hadron angular correlations and the
energy imbalance of reconstructed dijets, which are
observed to be strongly suppressed and modified, respectively, in central AA collisions compared
to a pp (vacuum) reference. Single particle measurements provide limited information on the initial parton energy and its
radiation. Jet reconstruction allows more direct access to the parton energies, which can be calculated using pQCD, by integrating over the hadronic degrees of freedom in a collinear and infrared safe way.
Jets are reconstructed by grouping the detected particles within a given angular region, e.g.\ a
cone with radius $R$. The interaction with the
medium can result in a broadening of the jet profile with respect to
vacuum fragmentation. In this case, for a given jet resolution parameter $R$ and a fixed
initial parton energy, the energy of the jet reconstructed in heavy-ion
collisions will be smaller than in vacuum. In the case where the
gluons are radiated inside the cone, the jet is expected to have a
softer fragmentation and a modified density profile compared
to jets in vacuum.

Jet measurements in heavy-ion collisions employ various approaches to correct for
background energy not associated with jet production and to suppress
the combinatorial, false jet yield induced by fluctuations of this
background, e.g.\ via energy or momentum thresholds for particles that
are used in the jet finding process. Every approach represents a
compromise between potential fragmentation biases in the jet
reconstruction and a better separation of the jet signal from the
background. 

In this article a measurement of the inclusive jet \pt{} spectrum in
\PbPb{} collisions at $\sNN=2.76$ TeV is reported in four
centrality intervals in the most central 80\% of the total hadronic cross-section. Jets are clustered from charged
tracks measured with the central barrel detectors in ALICE down to
momenta of 0.15\,GeV/$c$, which provides unique access to low
$\pt$ jet fragments at mid-rapidity at the LHC. Jets are measured with resolution
parameters $R=0.2$ and $R=0.3$ in the pseudo-rapidity interval
$-0.5 < \eta < 0.5$. The underlying event is subtracted
event-by-event for each measured jet. The jet spectrum is corrected for background fluctuations and
detector effects affecting the jet energy resolution and scale through an unfolding procedure.

The jet reconstruction strategy and the correction procedure for background from the underlying event is discussed in detail in Section \ref{sec:methods}. The results are presented in Section \ref{sec:results} and discussed in Section \ref{sec:discussion}. 
\enlargethispage{1cm}
\section{Data analysis and techniques}\label{sec:methods}

\subsection{Data Sample and Event Selection}
The data used for this analysis were recorded by
the ALICE detector \cite{Aamodt:2008zz} in the fall of 2010 during the first \PbPb{} run at
a collision energy of $\sNN = 2.76$~TeV.  The analysis presented here
uses minimum-bias events, which are selected online by requiring
a signal in at least two out of the following three detectors: the
forward VZERO counters (V0A and V0C) and the Silicon Pixel Detector
(SPD) \cite{Abelev:2013qoq}.  The VZERO counters are forward scintillator
detectors covering a pseudo-rapidity range of $2.8 < \eta < 5.1$ (V0A)
and $-3.7 < \eta < -1.7$ (V0C); the SPD is part of the
Inner Tracking System (ITS) described below. The minimum-bias
trigger is fully efficient in selecting hadronic events in \PbPb{} collisions. In
addition, an offline selection is applied in which the online trigger
is validated and remaining background events from beam-gas and
electromagnetic interactions are rejected.  To ensure a high tracking
efficiency for all considered events, the primary vertex was required
to be within 10 cm from the center of the detector along the beam axis and within 1 cm in the transverse plane.

The number of \PbPb{} events used in this analysis after event selection is $12.8$ million in a centrality range between 0 and 80\% most central of the total hadronic
cross section, corresponding to a total integrated luminosity of 2 $\mu \mathrm{b}^{-1}$. The event sample is divided in four centrality
intervals (0--10\%, 10--30\%, 30--50\%, and 50--80\%) based on the sum of VZERO amplitudes.  
A Glauber model is used to calculate the number of participating
nucleons $N_{\rom{part}}$ in the collisions, the number of binary
collisions $N_{\rom{coll}}$, and the nuclear overlap function
$T_{\rom{AA}}$ \cite{Abelev:2013qoq}. The resulting values and their uncertainties for the considered centrality intervals are given in Table
\ref{tab:NCollTAANPart}. 

\TabH{tab:NCollTAANPart}{
Average values of the number of participating nucleons
$N_{\rom{part}}$, number of binary collisions $N_{\rom{coll}}$, and the nuclear
overlap function $T_{\rom{AA}}$ for the centrality intervals used in
the jet analysis. Experimental uncertainties on the parameters of the
nuclear density profile used in the Glauber simulations and on
the interpolated nucleon-nucleon cross section
($\sigma_{\rom{inel}}^{\rom{NN}} = 64 \pm 5$ mb) are included in the
uncertainties. For details see \cite{Abelev:2013qoq}.
}
{cccc}
{
\toprule
Centrality & $\langle N_{\rom{part}}\rangle$ & $\langle N_{\rom{coll}}\rangle$ & $\langle T_{\rom{AA}}\rangle$\\
\midrule
0--10\%  & $356.0 \pm 3.6$ & $1500.5 \pm 165.0$ & $23.5 \pm 0.8$\\
10--30\% & $223.0 \pm 3.5$ & $738.8 \pm 75.3$ & $11.6 \pm 0.4$\\
30--50\% & $107.2 \pm 2.8$ & $245.6 \pm 23.3$ & $3.8 \pm 0.2$\\
50--80\% & $32.5 \pm 1.2$ & $45.9 \pm 4.6$ & $0.70 \pm 0.04$\\
\bottomrule
}

\subsection{Jet reconstruction}\label{sec:JetRec}

Jets were reconstructed using charged
tracks detected in the Time Projection Chamber (TPC) \cite{Alme:2010ke} and the Inner
Tracking System (ITS) \cite{Aamodt:2010aa} which cover the full azimuth and pseudo-rapidity
$|\eta|<0.9$.  For each track traversing the TPC, up to 159
independent space points are measured at radial distances from 85 cm to 247 cm.

The ITS consists of six cylindrical silicon layers with high granularity for precision tracking, with the inner layer at 3.9 cm from the center of the detector and the outer layer at 43 cm. 
The measured space points in the ITS and the TPC are combined to reconstruct the tracks of charged particles. The transverse momentum is calculated from the measured track curvature in the magnetic field of $B=0.5$ T.

The main track selection criteria
are a minimum number of points in the TPC, a $\chi^2$ cut on the
fit, and a cut on the difference between the 
parameters of the track fit using all the space points in ITS and TPC, and using only the TPC space points with the primary vertex position as an additional constraint. Tracks for which the total change in the
track parameters is more than $6\sigma$ ($\chisq > 36$) are rejected
from the sample resulting in a tracking efficiency loss of 8\% for low \pt{} tracks ($\pttrack<1$ \GeVc) and a few percent (1-2\%) for higher momentum tracks. 
For a large fraction (79\%) of the tracks used in the analysis, at least one point was found in one of the two inner pixel tracking layers (SPD) of the ITS. 
To improve the azimuthal uniformity of the selected tracks, tracks without SPD points were
also used in the analysis. For those tracks the momentum was determined
from a track fit constrained to the primary vertex, to guarantee good
momentum resolution.

The \pt{} resolution for tracks is estimated from the
track residuals of the momentum fit and does not vary significantly
with centrality. All track types have a relative transverse momentum
resolution of $\sigma(\pt)/\pt\simeq 1\%$ at 1 \GeVc. 
The resolution at $\pt=50$ \GeVc{} is $\sigma(\pt)/\pt\simeq
10\%$ for tracks that have at least three out of six reconstructed space points
in the ITS. For the remaining tracks (6\% of the track sample)
the resolution is $\sigma(\pt)/\pt\simeq 20\%$ at $50$
\GeVc{}. The track \pt{} resolution is verified by cosmic muon events and the width of of the invariant mass peaks of $K_{\rm S}^{\rm 0}$, $\Lambda$ and $\bar{\Lambda}$ \cite{Abelev:2012eq}.

The track finding efficiency at $\pt=0.15$ \GeVc{} is 60\% increasing to
$\sim 90$\% for $\pt \simeq 1.5$ \GeVc{} and then decreases to $\sim 86\%$ for  $\pt \geq 2.5$ \GeVc{}. In peripheral events the track
finding efficiency is $\sim 2$\% larger than in central collisions due
to the lower track multiplicity. 

Jets are reconstructed with the anti-\kT{} algorithm using the FastJet package
\cite{Cacciari2011, Cacciari2006} with
resolution parameters $R=0.2$ and $R=0.3$. Charged tracks with
$|\eta|<0.9$ and $\pt>0.15$ \GeVc{} are used as input for the jet
algorithm. The transverse momentum of the jets, \ptjetrawch, is
calculated with the boost-invariant \pt\ recombination scheme. The
area, $A$, for each jet is determined using the active area method as
implemented in FastJet \cite{Cacciari2008a}. So-called `ghost particles' with very small momentum ($\sim 10^{-100}$ GeV/$c$) are added to the event and
the number of ghost particles in a jet measures the area. Ghost
particles are uniformly generated over the tracking acceptance ($0<\varphi<2\pi$ and $|\eta|<0.9$), with 200 ghost particles per unit
area. Jets used in the analysis are required to have an area larger
than 0.07 for $R=0.2$ jets and 0.2 for $R=0.3$ jets. This selection
mostly removes low momentum jets with $\ptjetrawch < 20$ \GeVc.  Jets are selected to have $|\eta| < 0.5$, so that they are fully contained in the tracking acceptance. 
In addition, jets containing a track with a reconstructed $\pt>100$ \GeVc{} are rejected from the analysis, to avoid possible contributions from tracks with poor momentum resolution (the momentum resolution is 20\% for tracks with $\pt=100$ \GeVc).
This selection has negligible effect in the reported range of jet momenta.

\subsection{Background subtraction}\label{sec:BkgSub}
In \PbPb{} events, 
the large background consisting of particles from soft scattering processes as well as fragments from other jets, is subtracted using the procedure proposed in \cite{Cacciari2008,Cacciari2010}. The background is measured on an event-by-event basis by clustering all particles using the \kT-algorithm and determining the median of the transverse momentum density $\rho^i_{\rom{ch}}=\ptjetch^{i}/A^{i}$ of all clusters $i$ in the event, excluding the two leading clusters to limit the impact of the hard jet signal on the background estimate. 
The signal anti-\kt{} jets are then corrected for the average background contribution using the median $\rho_{\rom{ch}}$:
\begin{equation}
\label{eq:ptsub}
\ptjetch = \ptjetrawch - \rho_{\rom{ch}}\;A,
\end{equation}
with \ptjetch{} the background subtracted jet \pt{}, \ptjetrawch{} the uncorrected measured jet \pt{} and $A$ the area of the anti-\kt{} signal jet. The inclusive jet distribution is then corrected via unfolding to account for background fluctuations and detector effects.

As demonstrated in \cite{Abelev:2012ej} the measured background
density $\rho_{\rom{ch}}$ is directly related to the multiplicity and
average transverse momentum of the reconstructed charged particles.
Since it is based on the same collection of input particles used for
the signal jets, the quantity $\rho_{\rom{ch}}$ used in the analysis
intrinsically includes all detector effects, such as tracking
efficiency and momentum resolution. 
To enable comparisons with other experiments and generator studies, the corrected background momentum density is obtained using the Hit-Backspace-Once-More (HBOM) method proposed in \cite{Monk:2011pg}, i.e. by
repeatedly applying the parameterized detector response to the
measured heavy-ion events and extrapolating the measured $\rho$
to an ideal detector. The advantage of the method lies in the data-driven approach
where only the detector response is taken from simulation. This is of
particular importance when studying observables that are sensitive to
the a-priori unknown structure of the heavy-ion event and the
correlation between different regions in the event.
This procedure yields a corrected transverse
momentum density of $\rho_{\rm ch}^{\pt > 0.15} = 155.8 \pm 3.7$
\GeVc{} for the 10\% most central events, with a spread
$\sigma(\rho_{\rm ch}^{\pt > 0.15}) = 20.5 \pm 0.4$ \GeVc{} with no
significant dependence on the distance parameter $R$ employed in the
$\rho$ calculation. 

\subsection{Background fluctuations}\label{sec:BkgFluctuations}
All particles created in a collision are clustered into jets, but not all of them originate from hard processes. 
The distinction between jets
originating from a hard parton and soft clusters containing mostly
background particles (\textit{combinatorial jets}) is to some extent arbitrary and 
requires a pragmatic definition. At very high \pt, it is clear that
all jets originate from parton fragmentation processes, while
at low and intermediate \pt, clusters can be formed by including
fragments from multiple, independent parton scatterings or even from the soft hadronization. 

Jet clusters which originate from a hard scattering will contain
a large amount of uncorrelated, mostly soft, background particles. The
background subtraction procedure described in Section \ref{sec:BkgSub}, removes the
background energy \textit{on average}, but the background has large
region-to-region fluctuations in the event, both due to statistical
fluctuations of the particle number and momentum, and collective
phenomena like elliptic flow.

Combinatorial jets and background fluctuations are intimately
related: low energy jets, for example with a momentum below 5
\GeVc, are also subject to background fluctuations and appear at
relatively high \pt{} (well above 20 \GeVc). Such jets are mostly
background energy, and thus background fluctuations give rise to
combinatorial jets.

For the results reported in the next sections an unfolding procedure
is used to correct for background fluctuations. In this procedure, the
combinatorial jets will emerge at low \pt, while the spectrum
is only reported above a certain \pt{} cut-off, thus effectively
removing the combinatorial jets from the result.

\begin{figure}[tbh!f]
\includegraphics[width=0.5\textwidth]{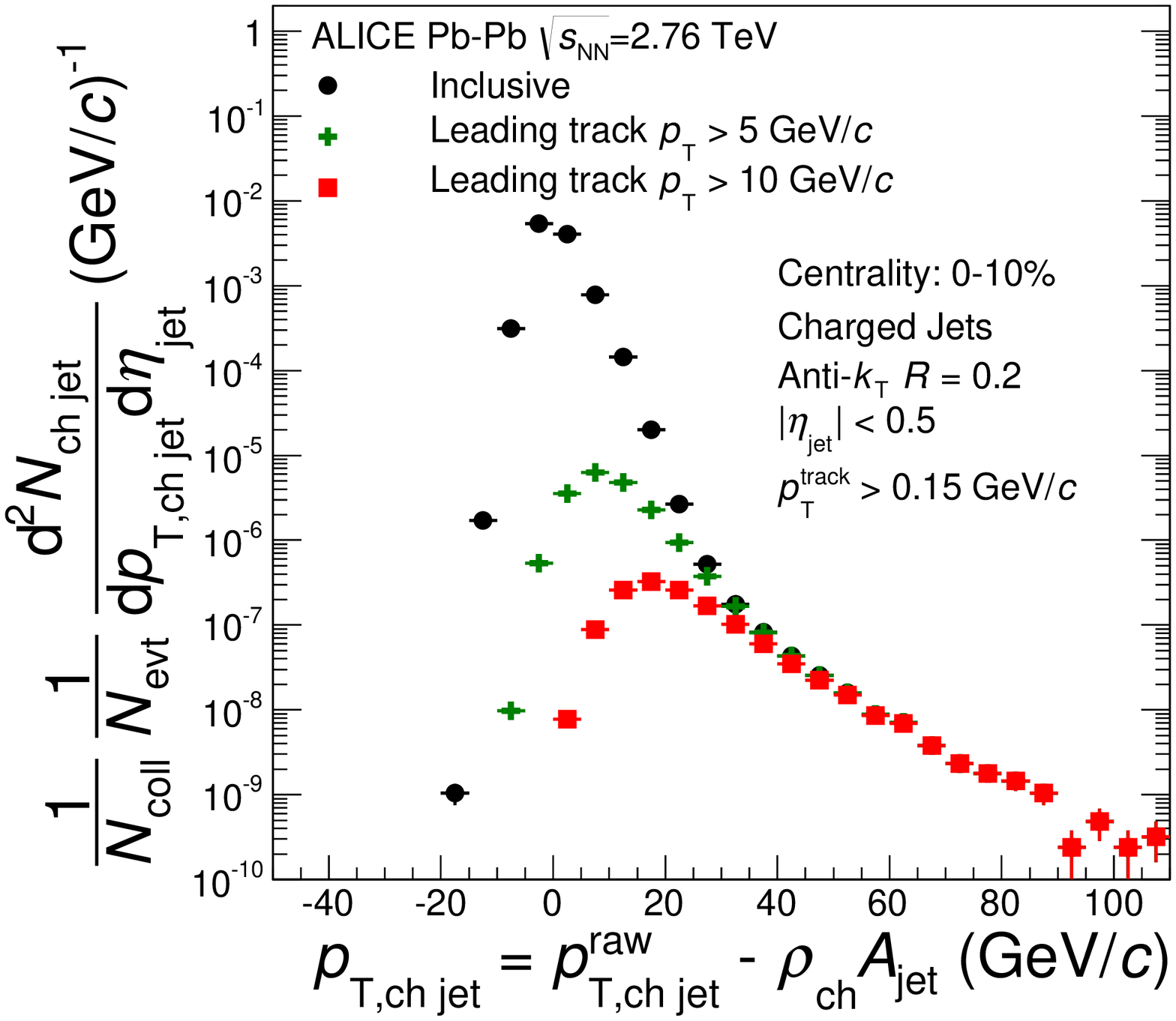}
\includegraphics[width=0.5\textwidth]{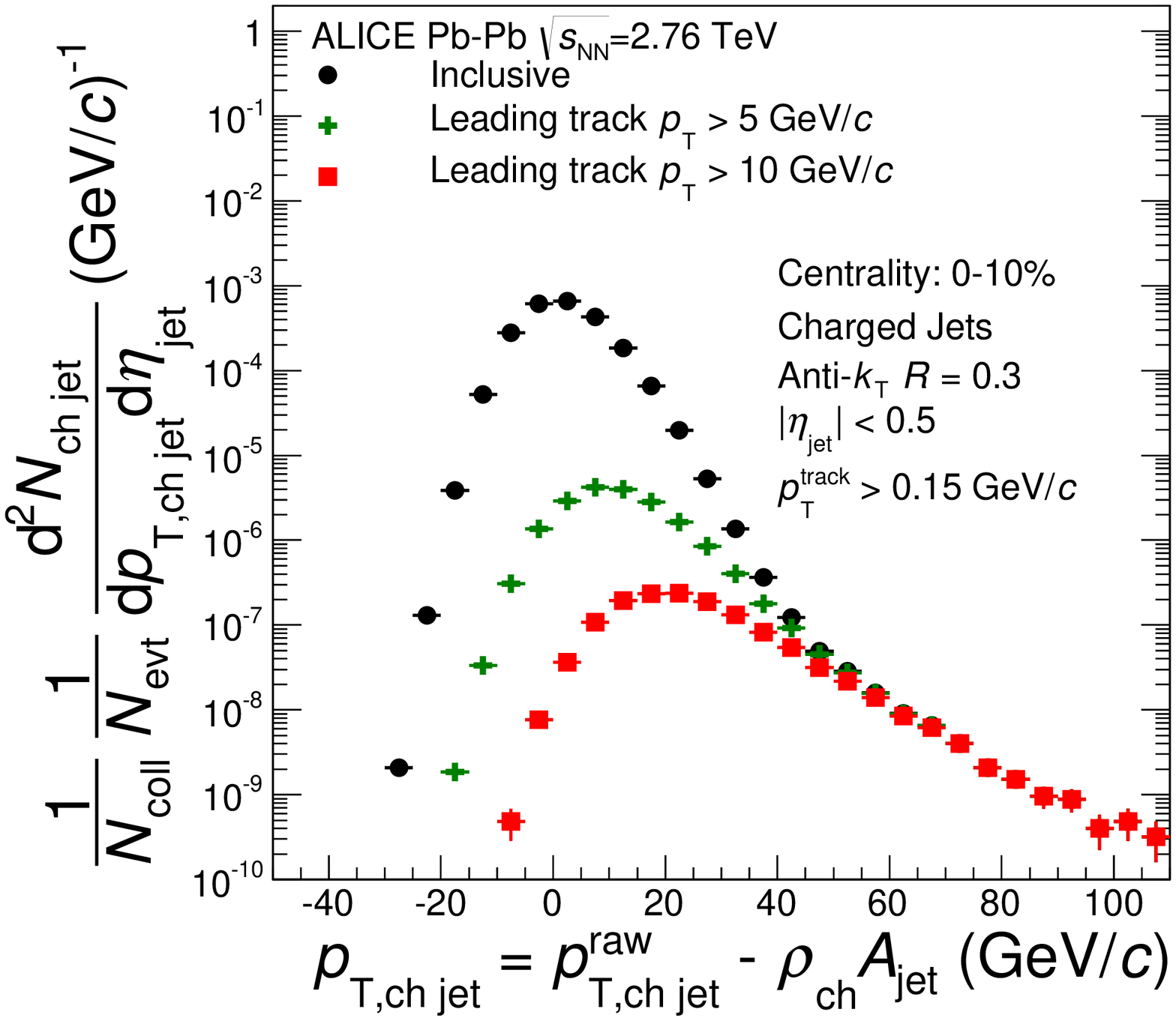}
\caption{\label{fig:rawspectra}Uncorrected jet spectra after
background subtraction, with radius parameters $R=0.2$ (left) and $R=0.3$ (right) in central
\PbPb{} events, without leading particle selection ($unbiased$, black circles) and
with at least one particle with $\pt > 5$ (green crosses) or $10$ \GeVc{} (red squares).}
\end{figure}

To illustrate the impact of combinatorial jets, Fig.~\ref{fig:rawspectra} shows uncorrected jet spectra after event-by-event
subtraction of the background following Eq.~\ref{eq:ptsub}. The black solid circles
show the result without further selection of the jets, which shows a
broad peak around $\ptjetch=0$ \GeVc. 
A large fraction of the combinatorial jets can be removed by selecting jets with a leading charged particle above a certain threshold \cite{deBarros:2012ws}.
The crosses and squares in Fig.~\ref{fig:rawspectra} show the jet spectra with a
leading charged particle above 5 and 10 \GeVc. It can be seen clearly 
that selecting jets by a leading high \pt{} particle reduces the background contribution for $\ptjetch<40$
\GeVc. However, this selection does not only reject combinatorial
jets, but also introduces a bias towards harder fragmentation.

In the following, unbiased and leading track biased jet spectra are reported. The systematic uncertainty arising from
the combinatorial jet correction by unfolding is smaller for the
biased spectra (for details, see Section \ref{sec:systematics}).

Fluctuations of the background are quantified by placing cones with
$R=0.2$ and $R=0.3$ at random locations within the acceptance of the
measured \PbPb{} events ($0<\varphi<2\pi$ and $|\eta_{RC}|<0.5$). The transverse momentum of charged particles
in the Randomly positioned Cone (RC) is summed and the difference $\deltapt =
\sum_{i}^{\rom{RC}}p_{\rm T,i} - \rho_{\rom{ch}} A$ is calculated, which represents the statistical
(region-to-region) fluctuations of the background. An alternative method to quantify the background fluctuations is also used in which 
high \pt{} probes are embedded into the \PbPb{} events \cite{Abelev:2012ej}. The events with embedded probes are clustered with
the anti-\kT{} jet finder and the transverse momentum \ptjetch{} containing the embedded probe in the heavy-ion
environment is compared to the embedded transverse momentum
$\pt^{\rom{probe}}$ by calculating the difference $\deltapt =
\ptjetrawch - \rho_{\rom{ch}} A - \pt^{\rom{probe}}$.

\begin{figure}[tbh!f]
\includegraphics[width=0.5\textwidth]{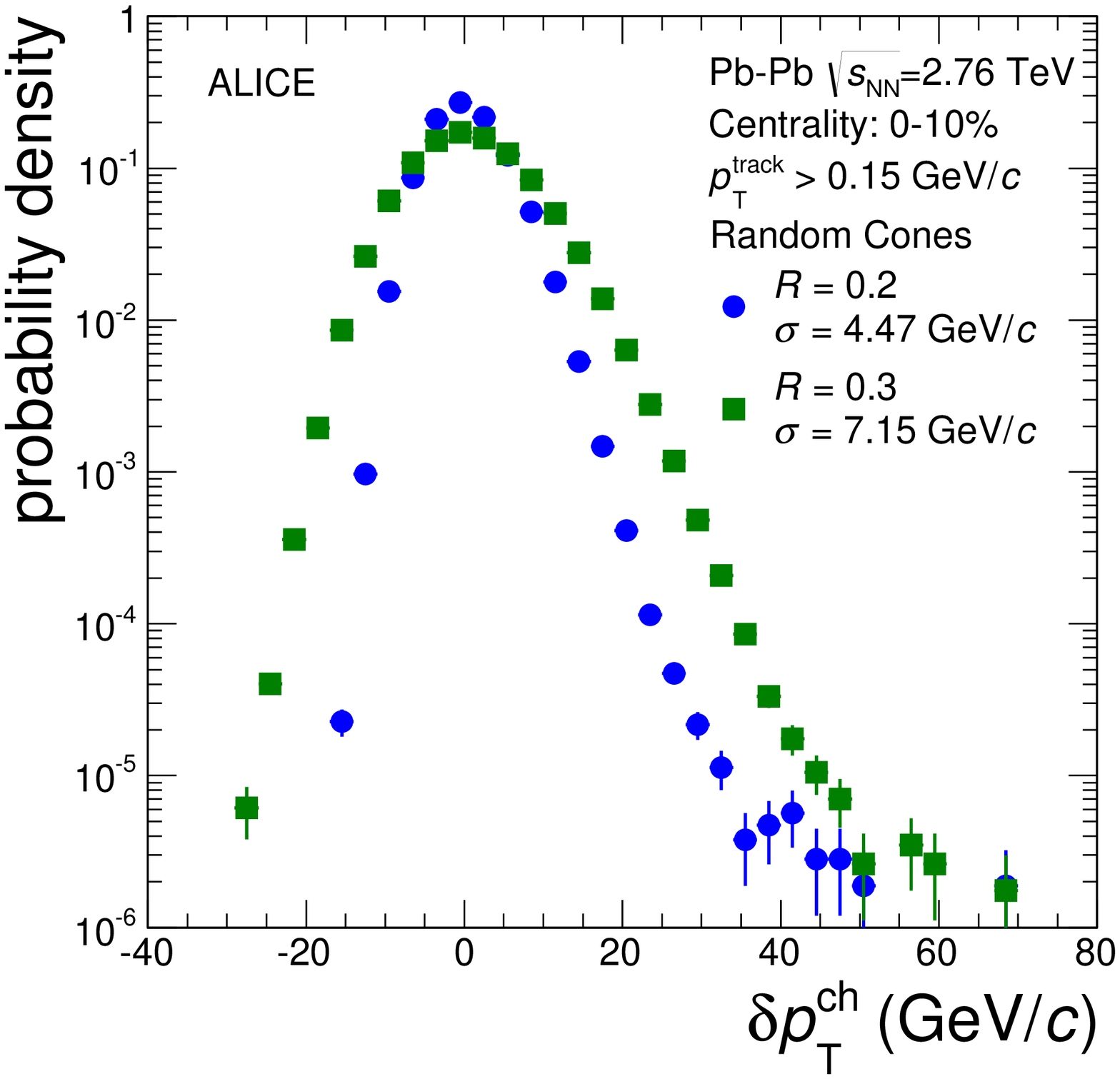}
\includegraphics[width=0.5\textwidth]{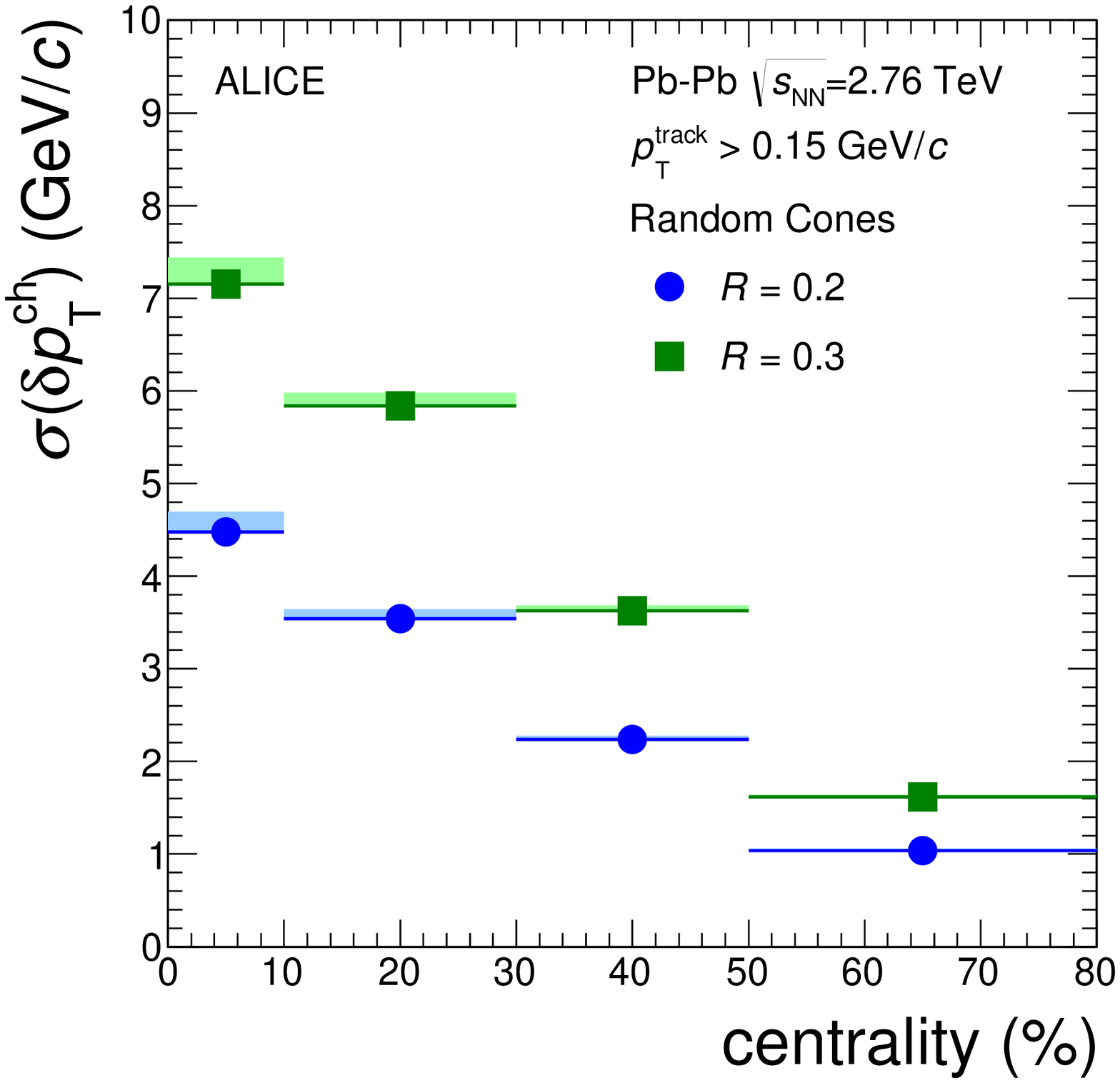}
\caption{\label{fig:BkgFluct}
Left: \deltapt{} distribution for jets with resolution parameter $R=0.2$
and $R=0.3$ measured with random cones in central collisions.  Right:
Width of the background fluctuation \deltapt{} distribution as a function of centrality for
cone radii $R=0.2$ and $R=0.3$. The shaded uncertainty bands indicate the difference between
the width of the \deltapt{} distribution from random cones and high
\pt{} probe embedding.  }
\end{figure}
The left panel of Fig.~\ref{fig:BkgFluct} shows the
\deltapt{} distribution from the 10\% most central events for the two
jet resolution parameters used in this analysis. The standard
deviation of the background fluctuations, $\sigma(\deltapt)$, is $4.47$ \GeVc{} for $R=0.2$ jets and $7.15$ \GeVc{} for $R=0.3$
jets (the statistical uncertainties are less than 4 \MeVc{} due to the large sample of random cones). The right panel of Fig.~\ref{fig:BkgFluct} shows the evolution of $\sigma(\deltapt)$ with centrality for the two jet resolution
parameters extracted with the random cones technique. The upper edge
of the shaded boxes indicates the $\sigma(\deltapt)$ obtained with
track embedding, where single tracks with $20<\pttrack<110$ \GeVc{}
were embedded in the heavy-ion events.  The small increase in the
standard deviation for more central events is due to the finite jet area resolution in the
embedding \cite{Abelev:2012ej}.

Due to the asymmetry of the \deltapt{} distribution, fluctuations
that increase the jet energy are more probable than fluctuations to lower jet energy. More importantly, the
steeply falling \pt-spectrum favours low-\pt{} jets with upward
fluctuations over downward fluctuations of high-\pt{} jets at a given
\pt.

Fluctuations of the background depend strongly on the multiplicity,
jet area (or radius), and minimum \pt{} of the measured particles \cite{Abelev:2012ej}. 
The analysis presented here is limited to $R=0.2$ and $R=0.3$ to avoid instabilities in the correction which are present for larger radii, see also Section \ref{sec:unfolding}.

The measured (uncorrected) \deltapt{} distributions are used directly to correct
the jet spectrum for background fluctuations. In addition, the magnitude of
background fluctuations also provides a potentially important
characteristic of the properties of the heavy-ion event and the
region-to-region variation of the transverse momentum density. For
this purpose, the measured values were corrected using the HBOM iterative procedure in the same way as for the background density $\rho$, 
\textit{i.e.} applying the parameterized detector
effects multiple times and extrapolating the fluctuations
to an ideal detector \cite{Monk:2011pg}. 
Since the correction is based on the properties of the measured
heavy-ion event, it takes into account all 
correlations in the event. The corrected width of the \deltapt{}
distribution is given in Table~\ref{tab:deltapt} for central
collisions and various cone radii. The FastJet package 
provides a measure of fluctuations, $\sigma_{\rom{FJ}}$, which is defined
from the distribution of individual jet momentum densities $\ptjetch^{i}/A^{i}$
such that 15.9\% of all clusters within an event satisfy $\ptjetch^{i}/A^{i} < \rho
- \sigma_{\rom{FJ}} \sqrt{A}$ \cite{Cacciari:2010te}.  This measure
corrects to first order the area dependence of fluctuations ($\sigma
\propto \sqrt{A}$), but is not sensitive to the tail of the
distribution. The $\sigma_{\rom{FJ}}$ obtained with different radius
parameters for the \kt{} jet finder and extrapolated to an ideal
detector for charged particles above $\pt > 0.15$ \GeVc{} is also
reported in Table~\ref{tab:deltapt}. It is
multiplied by $\sqrt{\pi R^{2}}$ to re-introduce
part of the area dependence, present in
$\sigma(\deltapt)$. The FastJet fluctuation
measures are reported to enable the comparison of fluctuations
in heavy ion reactions by standard jet reconstruction tools in models
and data.

\begin{table}[!ht]
\begin{center}
\renewcommand{\arraystretch}{1.3}
 \begin{tabular}{>{\small}l|>{\small}r>{\small}r|>{\small}r}
 &\multicolumn{2}{c}{\small{$\sigma(\deltapt)$}} & \multicolumn{1}{|c}{\small{$\sigma_{\rm FJ} \cdot \sqrt{\pi R^{2}}$}}  \\
  &  \multicolumn{1}{c}{Measured} &  \multicolumn{1}{c}{Corrected} &  \multicolumn{1}{|c}{Corrected} \\
\hline
$R = 0.2$ &$4.47 \pm 0.00$ \GeVc & $5.10 \pm 0.05$ \GeVc & $4.04 \pm 0.05$ \GeVc\\
$R = 0.3$ & $7.15 \pm 0.00$ \GeVc & $8.21 \pm 0.09$ \GeVc & $6.35 \pm 0.09$ \GeVc\\
$R = 0.4$ & $10.17 \pm 0.01$ \GeVc & $11.85 \pm 0.14$ \GeVc & $8.59 \pm 0.12$ \GeVc\\
\end{tabular} 
\end{center}
\caption{
\label{tab:deltapt}
Measured and corrected width of the $\deltapt$ distribution for
different cone radii in 10\% most central events for $\pttrack > 0.15$ \GeVc. In addition, the corrected fluctuation measure
from FastJet is provided, multiplied by $\sqrt{\pi R^{2}}$ to take into account the expected area dependence of the fluctuations. The values for $R=0.4$ are given for comparison with \cite{Abelev:2012ej}.}
\end{table}
\subsection{Detector effects}\label{sec:DetectorEffects}

The jet response in the ALICE detector is evaluated using simulations
with the PYTHIA6 \cite{Sjostrand2006} event generator and GEANT3
\cite{Brun1994} for detector response, using the same reconstruction software settings that are used for the reconstruction of \PbPb{} events. 
The effect of the high track density in \mbox{\PbPb{}} events on the tracking
efficiency was studied using HIJING \cite{Wang:1991hta} events with
the GEANT3 detector simulation. It is found that the tracking
efficiency is $\sim$2\% lower in central \PbPb{} collisions than
in peripheral collisions and \pp{} collisions. This additional
centrality-dependent inefficiency was introduced to the PYTHIA events by a random rejection
of tracks.

The jet response is determined on a
jet-by-jet basis by comparing jets before (particle level jets) and after detector simulation (detector level jets), that are geometrically matched.
Particle level jets are clustered from primary charged particles produced by the event generator. Primary charged particles include all prompt charged particles produced in the collision, including the products of strong and electromagnetic decays, but excluding weak decays of strange hadrons.
In this analysis the detector to particle level correction is based on the Perugia-0 tune
\cite{Skands:2010ak} of PYTHIA6. It was verified that the
simulated detector response for jets is largely independent of the generator tune by comparing to the jet response obtained with the Perugia-2010 and 2011 tune \cite{tagkey2013262,Vajzer:2013gla}. The contribution from weak decay products to the track sample is small due to the track selection requirements and low material budget ($11.5\% \pm 0.5\%$ $X_{\rm 0}$ in the central tracking systems \cite{Abelev:2012cn}). The remaining contamination is included in the response matrix.
No correction for hadronization effects was applied since the relation between parton level jet and particle level jet in heavy-ion collisions is not well-defined.

The detector effects that influence the jet energy scale and
resolution are the charged particle tracking efficiency and the
transverse momentum resolution, with the tracking
efficiency being the dominant contributor. The finite \pt{} resolution
of reconstructed charged tracks has a small effect on the jet energy
resolution since the majority of the constituents of a jet are of
moderate \pt{} where the tracking momentum resolution is
good. In addition, since the transverse
momentum of the jet is the sum of the transverse momentum of
independently measured tracks, the relative momentum resolution is in
general better than that of individual tracks.

\begin{figure}[tbh!f]
\includegraphics[width=0.5\textwidth]{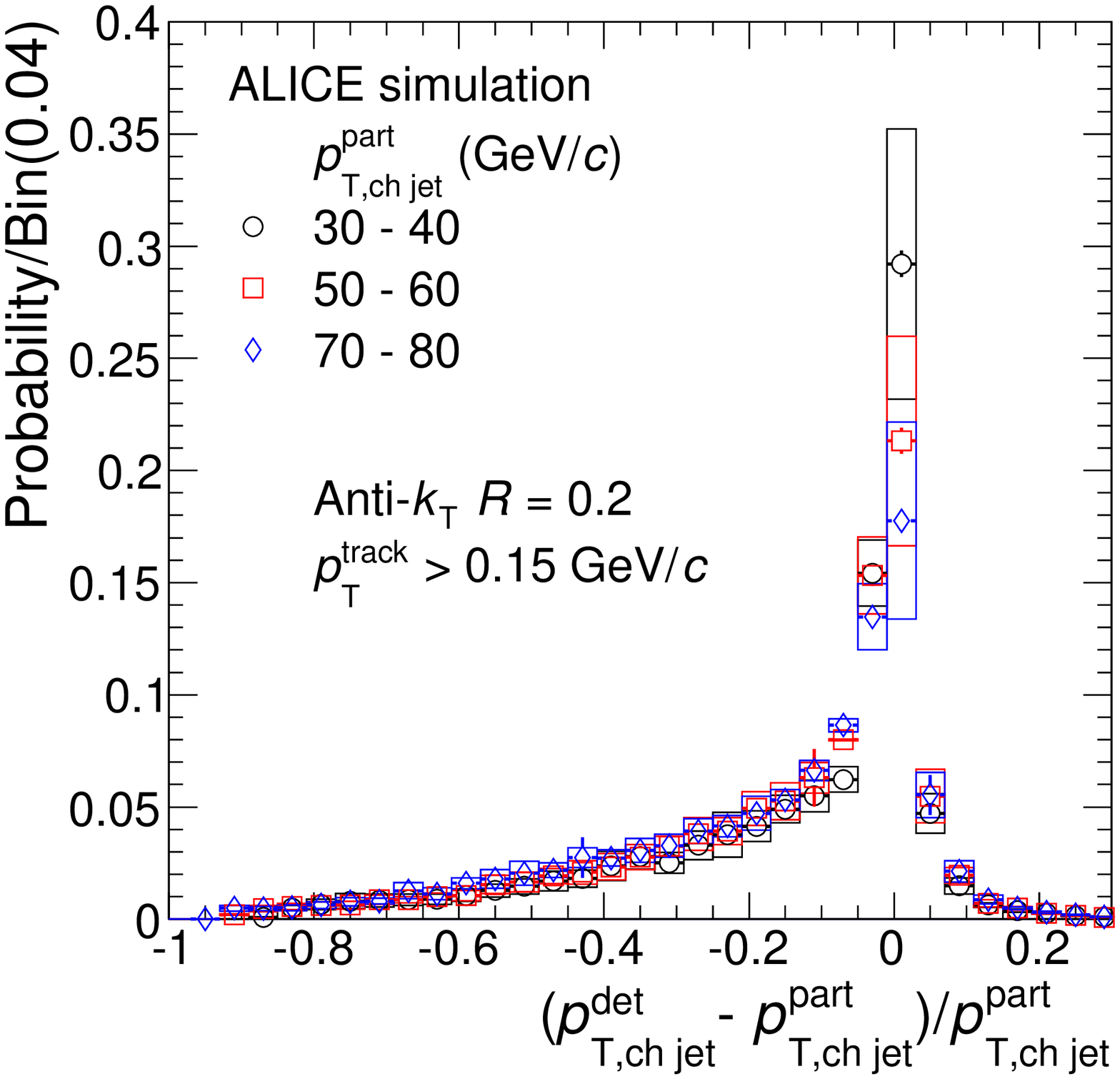}
\includegraphics[width=0.5\textwidth]{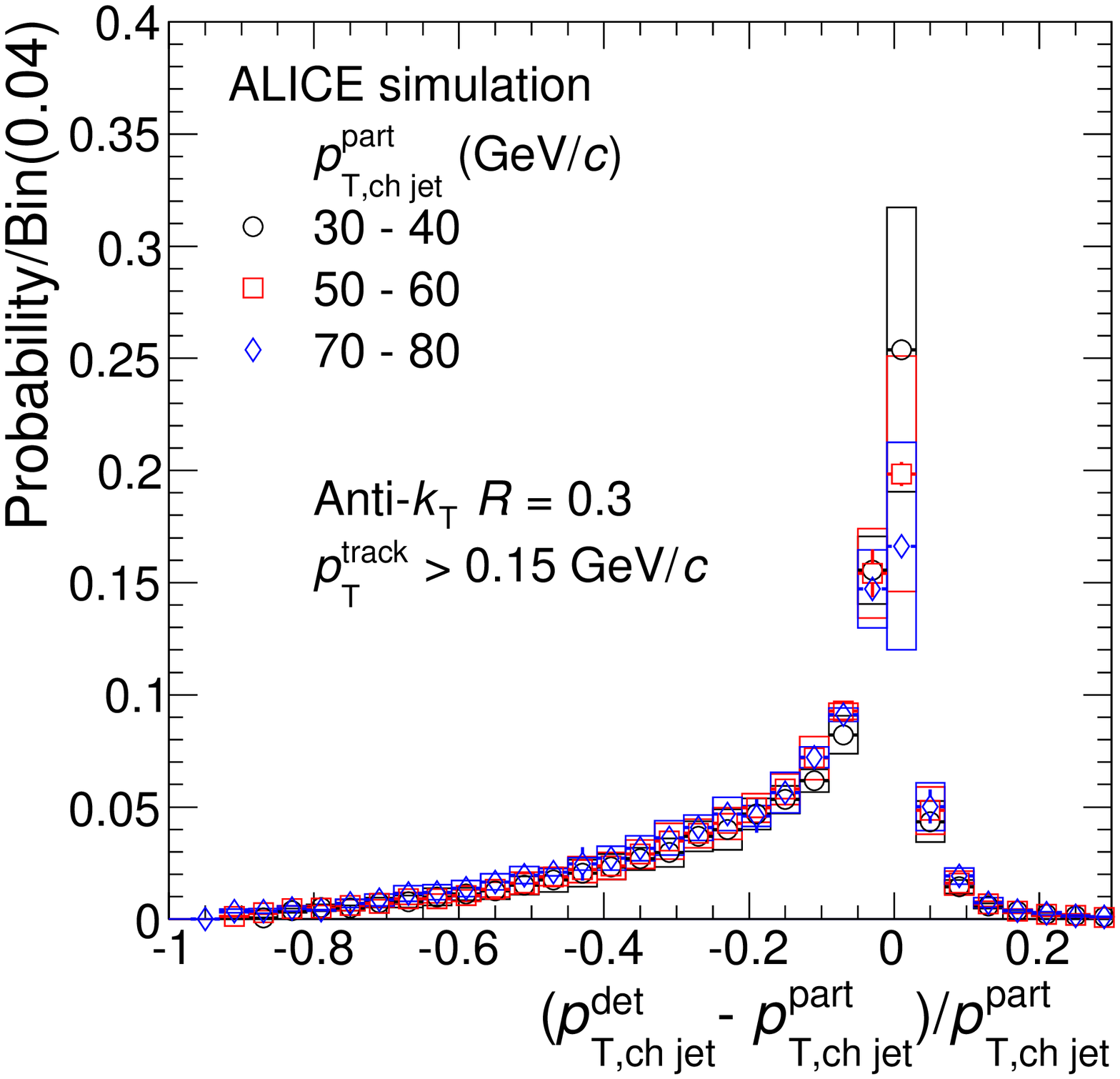}
\caption{\label{fig:jet_response}Distributions of relative transverse momentum difference between detector and particle level \mbox{anti-\kT{}} jets  with $R=0.2$ and  $R=0.3$  and several ranges of jet transverse momentum at particle level. The distributions correspond to the 10\% most central events. Events were generated using PYTHIA with the standard ALICE detector response simulation using GEANT3 and the data reconstruction algorithms and settings used for \PbPb{} events.  The dominant systematic uncertainty is the uncertainty on tracking efficiency.} 
\end{figure}
Figure \ref{fig:jet_response} shows the probability distribution of
the relative transverse momentum difference between the detector and
particle level jets with resolution parameters $R=0.2$ and $R=0.3$ in
three different intervals of the transverse momentum of the particle level jet \ptjetgen. The most probable detector
level \ptjetrec{} is very close to the particle level jet \ptjetgen{}
in all cases.
The average momentum of the detector
level jet is lower than the particle level momentum, because of the
average inefficiency of 10-20\% in the charged particle
reconstruction. Momentum resolution effects and under-subtraction of the background (back reaction) can
cause a detector level jet to have a higher momentum. The
momentum difference distribution is highly asymmetric and
cannot be described by a Gaussian distribution.

\begin{figure}[tbh!f]
\includegraphics[width=0.5\textwidth]{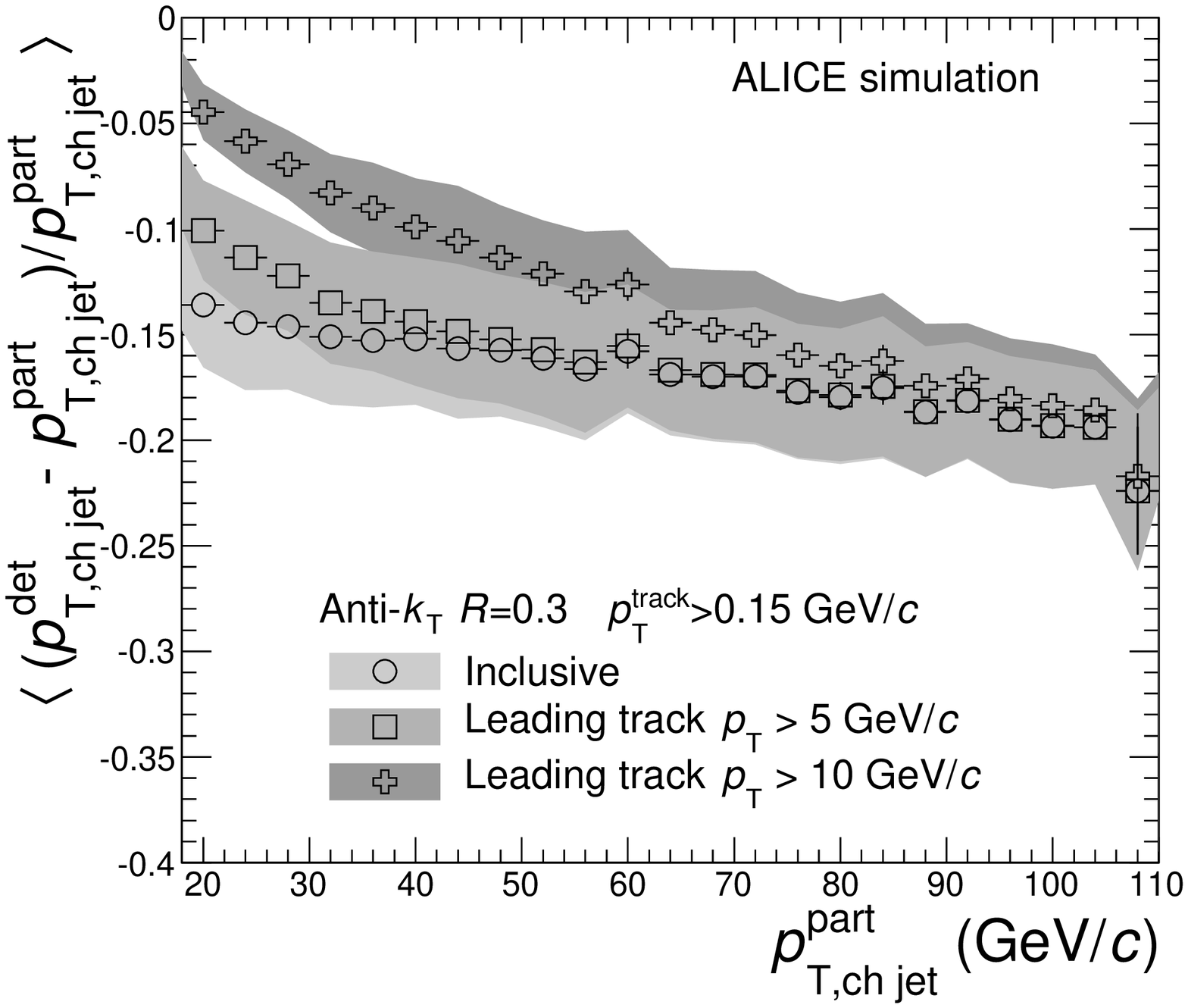}
\includegraphics[width=0.5\textwidth]{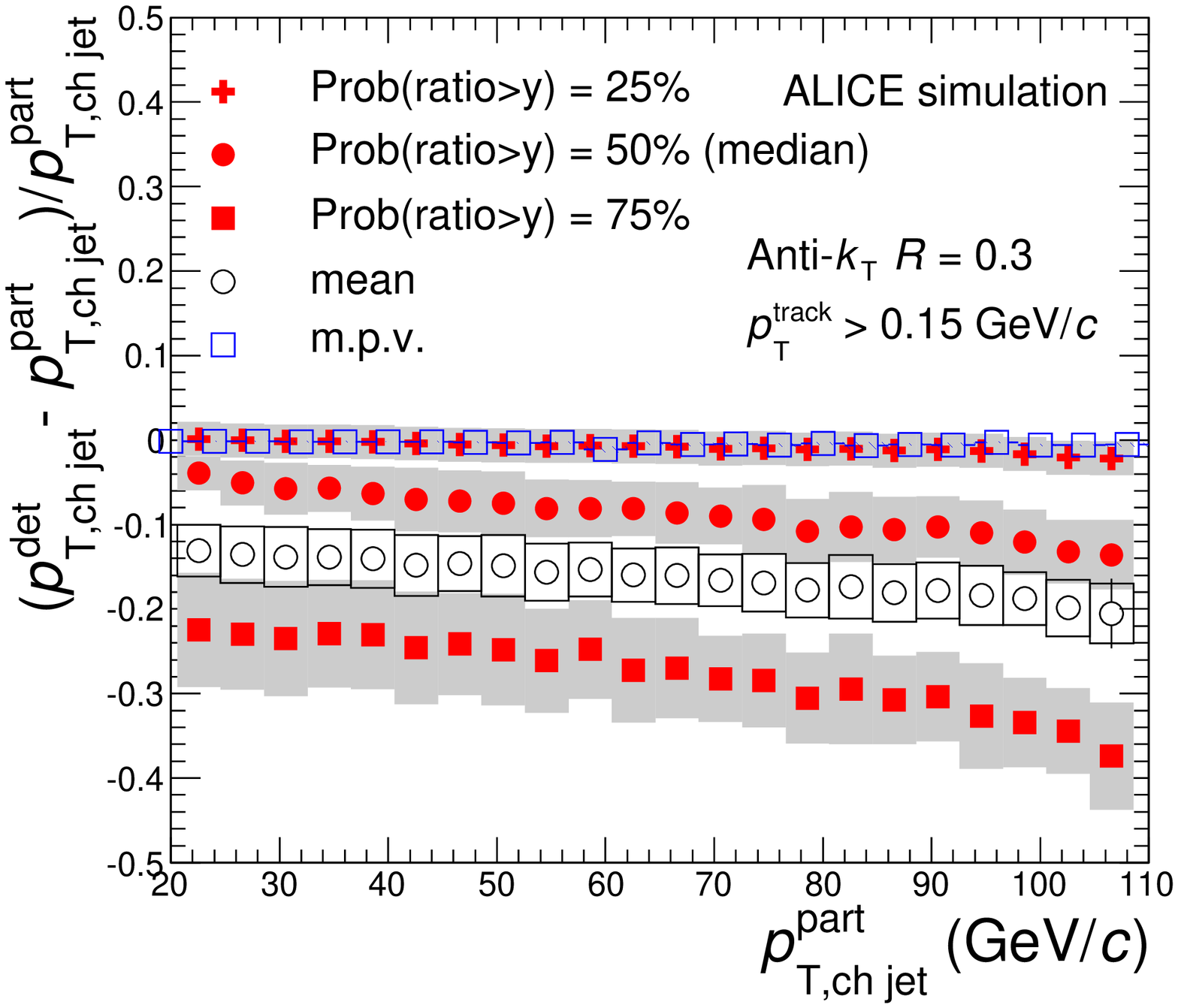}
\caption{\label{fig:jer}Jet detector response for jet finding resolution parameter $R=0.3$ for the 10\% most central events. Data points extracted from event and full detector simulation. Systematic uncertainty originates from the uncertainty on the tracking efficiency. Left: mean of the jet response for charged jets with $R=0.3$. See text for details. Right: Mean, most probably value and quartiles of the jet response as a function of jet momentum.}
\end{figure}

\begin{figure}[tbh!f]
\includegraphics[width=0.5\textwidth]{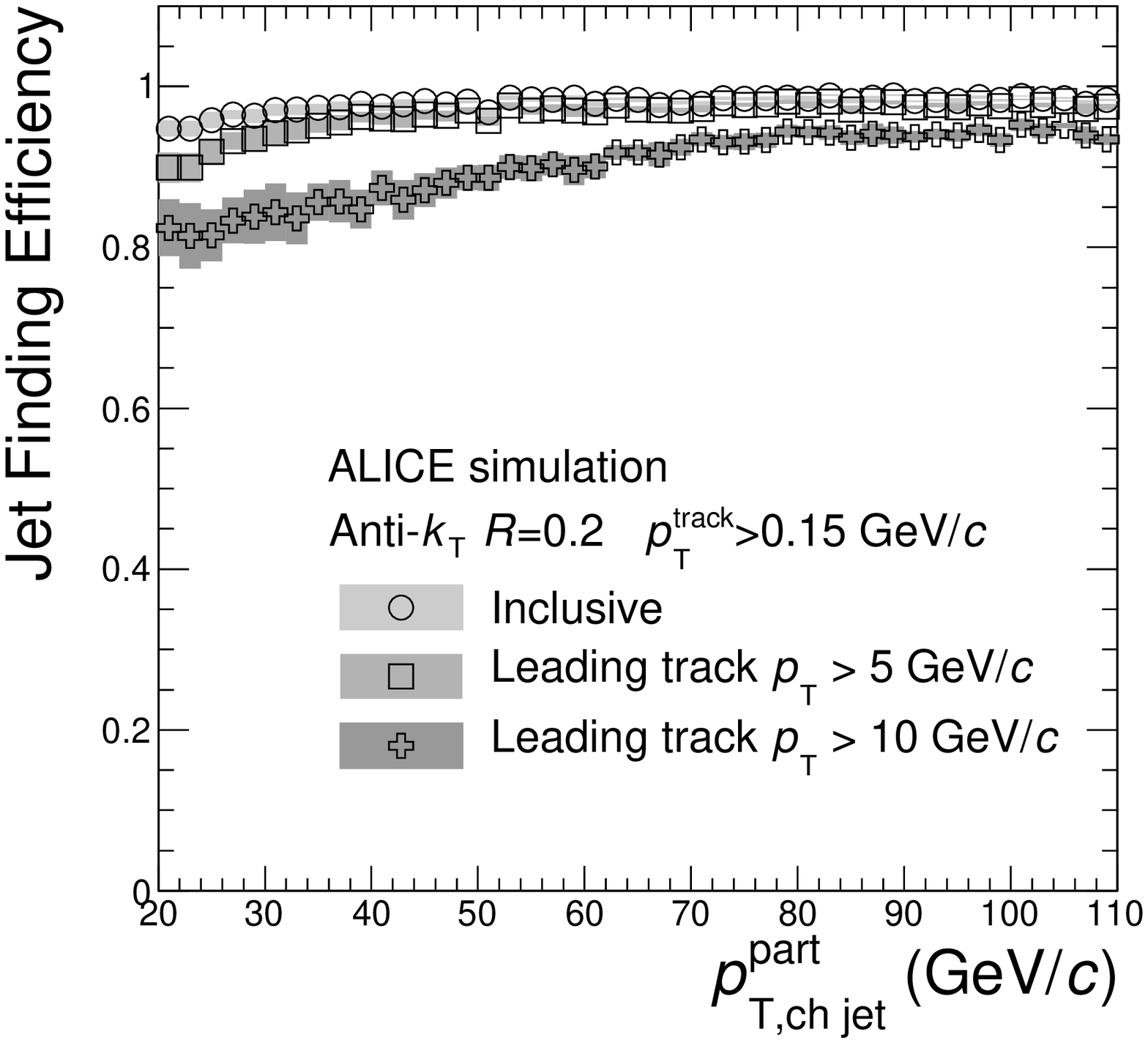}
\includegraphics[width=0.5\textwidth]{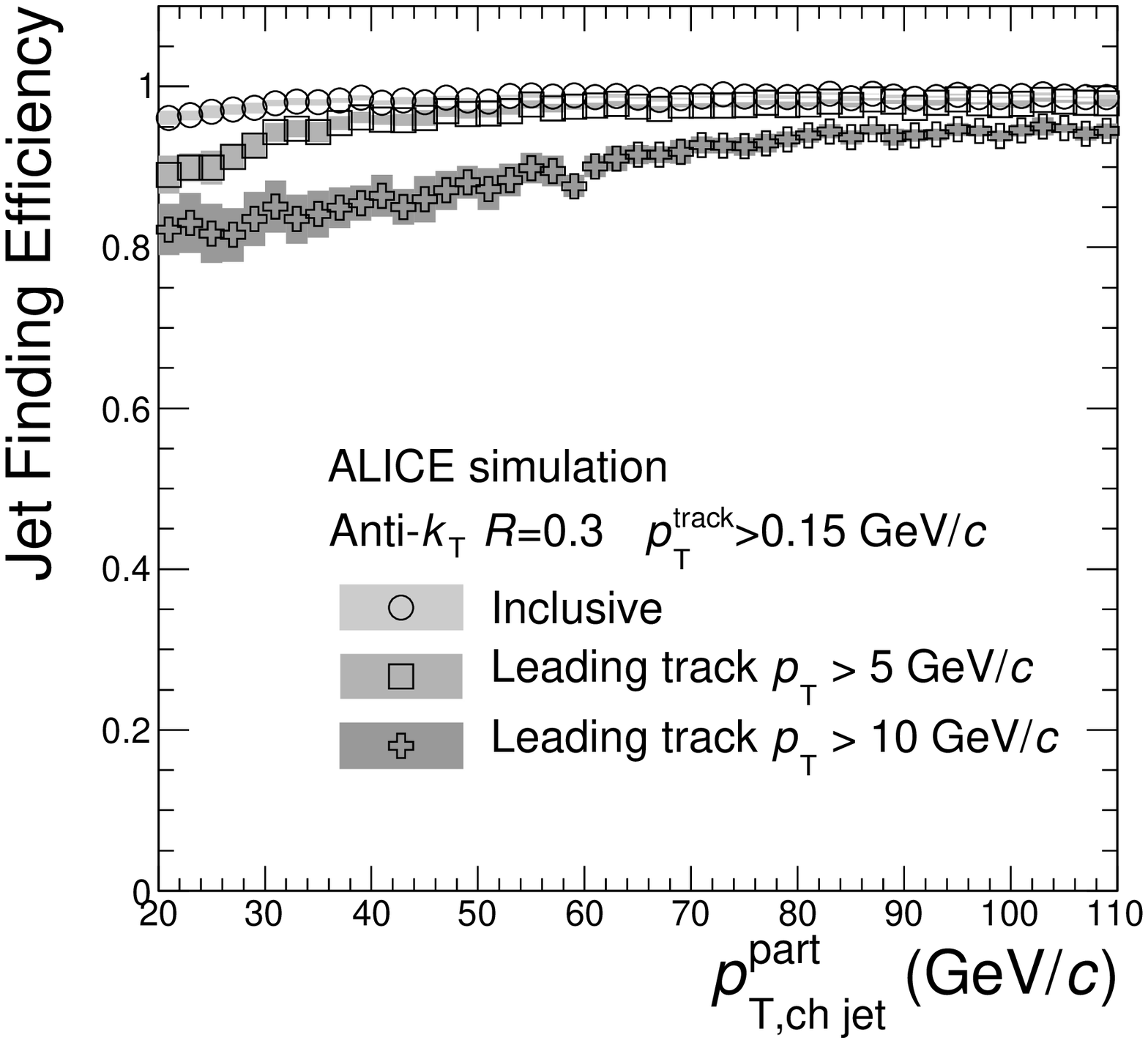}
\caption{\label{fig:JetFindingEfficiencyLeadTrack}
Jet-finding efficiency for inclusive unbiased and leading track biased jets extracted from event and detector simulation for the 10\% most central events. Left panel: $R=0.2$. Right panel: $R=0.3$.
}
\end{figure}

To characterize the detector response, the mean of the relative
difference between \ptjetrec{} and \ptjetgen{} as a function of the jet
momentum at particle level is shown in Fig.~\ref{fig:jer}. For unbiased jets the reconstructed jet momentum is on average
14--19\% lower than the generated momentum, in the range
$\ptjetgen=20-100$ \GeVc, with a weak \pt-dependence. The mean of
the jet response is also shown for leading track biased jets with $\ptleadtrack>5$ and $10$ \GeVc. Those jets whose
leading track is not reconstructed in the detector are rejected from
the sample. This results in an improved jet energy resolution at low jet
\pt{} while the jet finding efficiency is decreased, as shown in
Fig.~\ref{fig:JetFindingEfficiencyLeadTrack}.

To give more details on the detector response to jets, the most probable value of the relative difference between \ptjetgen\ and \ptjetrec\ is shown as a function of \ptjetgen{} in the right panel of Fig.~\ref{fig:jer}.
The most probable value is determined as the mean of a Gaussian function fitted to the peak region, $-0.03< (\ptjetrec-\ptjetgen)/\ptjetgen <0.03$. The most probable value of the detector level \pt\ is within 0.5\% of \ptjetgen over the entire \pt{} range.

The right panel in Fig.~\ref{fig:jer} also shows the boundaries at 25\%, 50\% or 75\% of the response distribution for jets with $R=0.3$, integrating from the right $\ptjetrec \rightarrow \infty$.
Approximately 25\% of the detector level jets has a larger reconstructed jet momentum than generated. The 50\% percentile (median) correction is 5\% at $\ptjetgen=20$ \GeVc\ and increases to 14\% at $\ptjetgen=100$ \GeVc. For 75\% of the jet population the correction for detector effects is smaller than 22\% at low $\ptjet\approx20\,\GeVc$ and 30\% at high $\ptjet\approx100\,\GeVc$.

The jet-finding efficiency is obtained by taking the ratio between the spectra of the particle level jets which have a
detector level partner, and all particle level jets. In case of jets
biased by a high \pt{} constituent, the numerator consists of jets
fulfilling the high \pt{} track requirement on detector level and the
denominator are all particle level jets with a high \pt{} generated
particle.  Figure \ref{fig:JetFindingEfficiencyLeadTrack} shows the
jet-finding efficiency for the unbiased sample, which is
unity at high \pt{} and reduces to 95\% at $\ptjetgen=20$ \GeVc{} due
to migration of the jet axis outside the $\eta$ acceptance. The jet-finding
efficiency for jets with radii of $R=0.2$ and $R=0.3$ differs by
a few per cent at low \pt\ and is the same at high \pt. 
In general the jet-finding efficiency is $\sim$1\% higher in pp compared to
\PbPb{} without a \pt{} dependence for $\ptjetgen > 20$ \GeVc. 
For leading track biased jets, 
the jet-finding efficiency is reduced and reaches 90\% at
$\ptjetgen \approx 25$ \GeVc\ for $\ptleadtrack>5$ \GeVc\ and at
$\ptjetgen \approx 60$ \GeVc\ for $\ptleadtrack>10$ \GeVc, which is consistent with the charged particle tracking efficiency.

\subsection{Unfolding}\label{sec:unfolding}

Both background fluctuations and detector effects lead to smearing of the measured jet momentum in heavy ion collisions. These effects can be corrected for using deconvolution, or {\it unfolding} procedures \cite{D'Agostini:1994zf,Blobel:2002pu,Schmelling1994400}. 
The background fluctuations and detector effects partially compensate: an upward energy shift is more likely due to background fluctuations while detector effects mainly induce a shift to lower \pt.

The relation between the measured spectrum $\mathbf{M}_{m}$ and the `true' jet spectrum $\mathbf{T}_{t}$ is
\begin{equation}\label{eq:matrixProblem}
\mathbf{M}_{m} = \mathbf{R}^{\mathrm{tot}}_{m,t} \cdot \mathbf{T}_{t} = \mathbf{R}^{\mathrm{bkg}}_{m,d} \cdot \mathbf{R}^{\mathrm{det}}_{d,t} \cdot \mathbf{T}_{t},
\end{equation}
where $\mathbf{R}^{\rom{det}}_{d,t}$ is the response matrix for detector
effects (including efficiencies), $\mathbf{R}^{\rom{bkg}}_{m,d}$ is the
response matrix for background fluctuations, and $\mathbf{R}^{\rom{tot}}_{m,t}=\mathbf{R}^{\rom{bkg}}_{m,d} \cdot \mathbf{R}^{\rom{det}}_{d,t}$ is the total response matrix for the combined effects of background fluctuations and detector effects. The subscripts $m,d,t$ are indices indicating the bin number.

The response for background
fluctuations is extracted with the data-driven method described in Section \ref{sec:BkgFluctuations} and the response for
detector effects is obtained from detector simulations as described in
Section \ref{sec:DetectorEffects}. The response matrices are combined into an overall response matrix $\mathbf{R}^{\rom{tot}}_{m,t}$. It was verified that correcting for detector effects and background fluctuations in two separate unfolding steps yields the same unfolded jet spectrum.

\begin{figure}[tbh!f]
\includegraphics[width=0.5\textwidth]{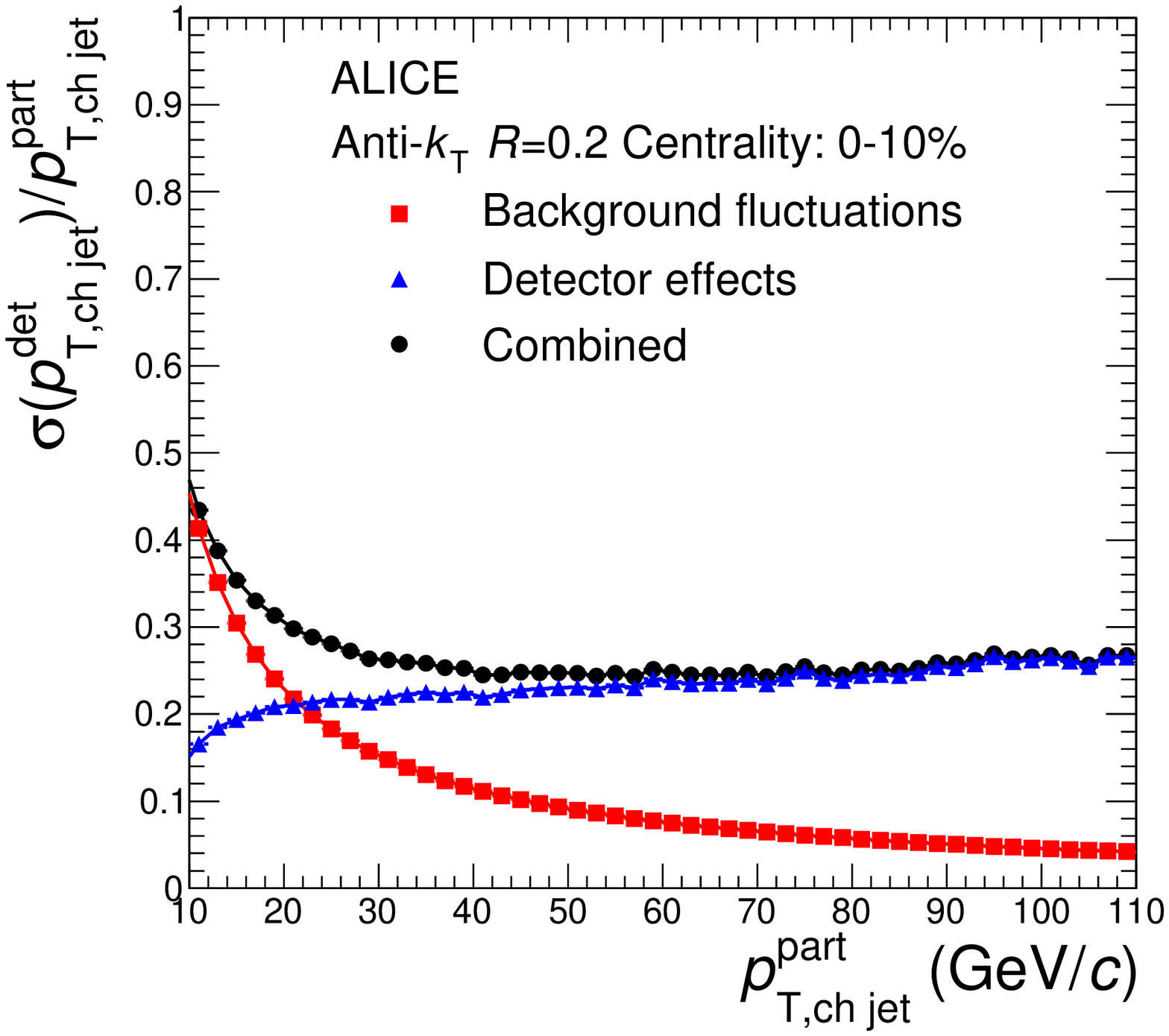}
\includegraphics[width=0.5\textwidth]{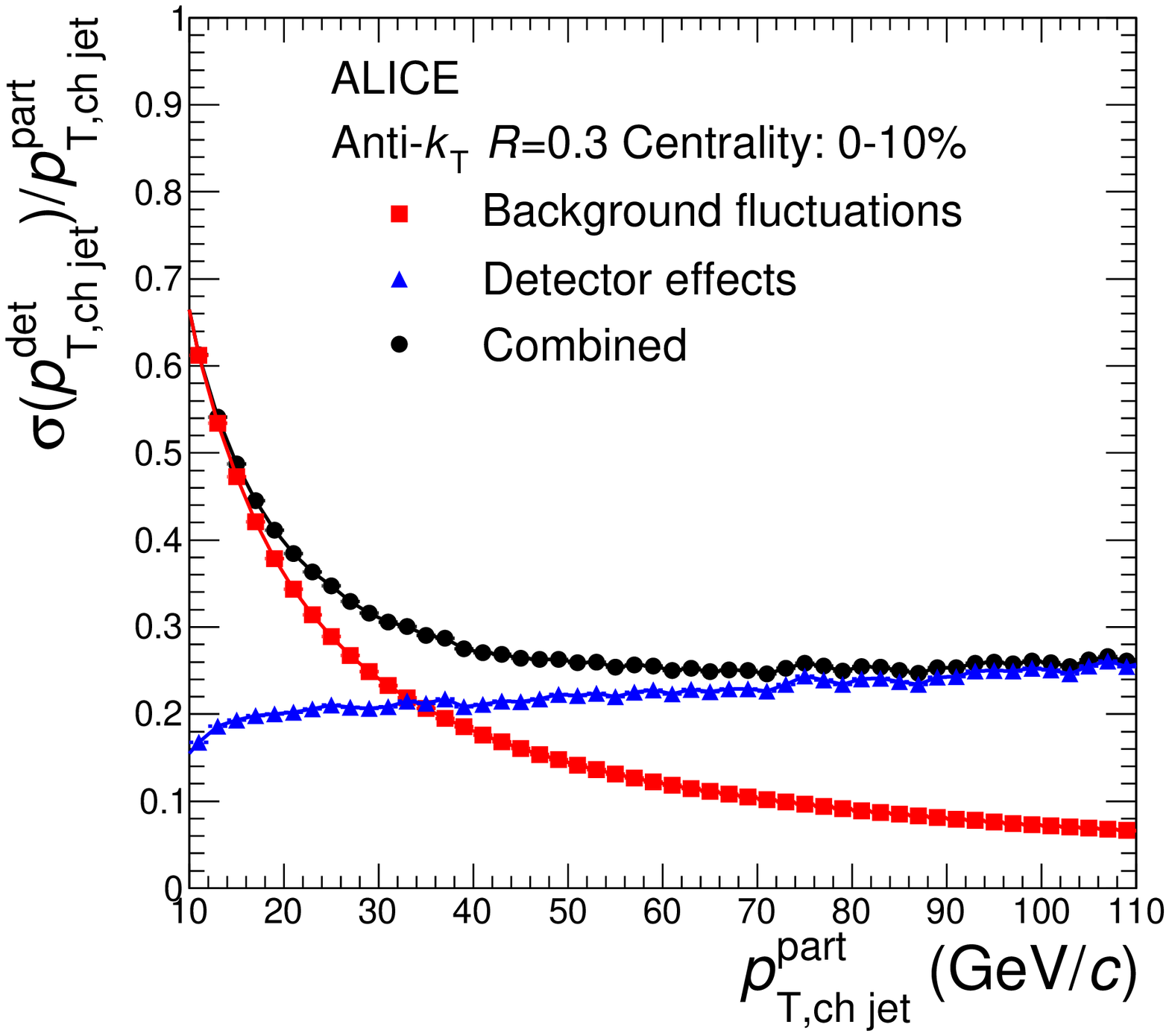}
\caption{\label{fig:combined_resp}
Combined jet response for charged jets for the two resolution parameters considered, including background fluctuations and detector effects for 0-10\% central \PbPb{} events. Left panel: $R=0.2$. Right panel: $R=0.3$.
}
\end{figure}

Figure \ref{fig:combined_resp} shows the width of the combined
response $\sigma(\ptjetrec)/\ptjetgen$ as a function of \ptjetgen. It can be observed that the
dominant correction at low momenta originates from the background
fluctuations while at high \pt{} the detector effects dominate.

Inverting Eq. \ref{eq:matrixProblem} to obtain the true spectrum
from the measured spectrum requires some care: calculating the inverse
of the response matrix leads to solutions for the true jet spectrum that has large unphysical
bin-to-bin-fluctuations. To suppress these fluctuations, unfolding
algorithms implement regularisation procedures, which impose a
smoothness criterion on the final result. There is some freedom in the
choice of regularisation procedure, which leads to an additional
systematic uncertainty on the unfolded spectrum for the final result.

Three unfolding algorithms with different regularisation procedures
were tested: the $\chi^2$ method with a log-log-regularisation (see
Appendix \ref{app:UnfMethods}), the (generalized) Singular Value Decomposition (SVD) method as
implemented in RooUnfold, and the Bayesian method
\cite{Blobel1984,Blobel:2002pu,dagostini:2003,Hocker:1995kb,Adye:2011gm}. 
It was found in a closure test with a thermal background model that the Bayesian method does not converge properly for this case, 
while the other two methods give similar results.
The covariance matrix $\rom{cov}(x,y)$ for the unfolded result is calculated by propagating the measurement errors in the unfolding and/or using Monte Carlo variations of the input spectra \cite{Adye:2011gm}. The quality of the unfolded result is evaluated by inspecting the Pearson coefficients $\rho(x,y) = \frac{\rom{cov}(x,y)}{\sigma_{x}\sigma_{y}}$. A large (anti-)correlation between neighboring bins indicates that the regularisation is too strong or too weak. The statistical uncertainties on the unfolded data points are the square root of the diagonal elements of the covariance matrix of the unfolded spectrum.

\subsubsection{Unfolding strategy -- \pt{} ranges}\label{sec:UnfStrategy}
There are two relevant kinematic ranges in the unfolding strategy applied in this analysis: 
the \pt-range of the measured spectrum and the \pt-range of the unfolded
spectrum, which may be different.  A minimum \pt{} cut-off
on the measured jet spectrum is introduced to suppress jet candidates,
which are dominated by background fluctuations, including
combinatorial jets, while the unfolded spectrum starts at the lowest
possible \pt, $\pt^{\rom{unfolded}}>0$ \GeVc.

The minimum \pt{} cut-off ($\ptminmeas$) on the measured spectrum removes a large fraction of combinatorial jets, which makes the unfolding procedure more stable.
Feed-in from true jets with $\pt<\ptminmeas$ into the region used for unfolding is accounted for by extending  the unfolded spectrum to $\ptjetch=0$ \GeVc. The feed-in from low \pt{} true jets is a significant effect since the spectrum falls steeply with \ptjetch. 
Combinatorial jets still present in the measurement after applying the kinematical selections are transferred in the unfolding procedure to the region below \ptminmeas. 
Feed-in from jets with \ptjetch{} larger than the maximum measured \ptjetch{} is also included by extending the reach of the unfolded spectrum to $\ptjetch = 250$ \GeVc.
The optimal value of the minimum \pt{} cut-off has been studied using
the jet background model described in \cite{deBarros:2012ws} and within
simpler set-up in which a jet spectrum is folded
with the measured background fluctuations. 
Stable unfolding is obtained with a minimum
\pt{} cut-off of at least five times the width of
the \deltapt{}-distribution $\sigma(\deltapt)$. For the most central
collisions and $R=0.3$, this means that the spectrum is
reported for $\ptjetch > 40$ \GeVc. In addition, the maximum \pt{}
cut-off is driven by the available statistics. The present data set
allows for a measurement of $\ptjetch<110$ \GeVc{} in central events
and $\ptjetch<90$ \GeVc{} in peripheral events.
In case of leading track biased jets,
the unfolding is more stable since the correction for combinatorial jets is reduced.

\subsection{Systematic uncertainties}\label{sec:systematics}

The systematic uncertainties on the results were evaluated by varying a number of key assumptions in the correction procedure and by using different unfolding methods. The different tests and the resulting systematic uncertainties are discussed in the following subsections, and summerized in Table \ref{tab:SystematicsAllSources}.

\subsubsection{Unfolding and regularisation uncertainties}

The uncertainties from the regularisation and the unfolding procedure were evaluated by changing the regularisation strength $\beta$ in the
$\chi^2$-method and by comparing the results from the $\chi^2$ method and
the generalised SVD method. Both variations give an uncertainty on the applied regularisation. Therefore, the uncertainties were taken to be the
maximum deviation from both studies. The SVD method also makes use of a prior, which was varied. This has a negligible effect on the result.

\paragraph{Regularisation strength $\beta$}
The regularisation strength $\beta$ (see
Eq. \ref{eq:chi2functionGenPen}) is varied from a value where fluctuating solutions dominate to the point
where the unfolding becomes over-constrained.  The main effect of varying $\beta$ is that
the unfolded jet spectrum changes shape. With increasing
regularisation, the unfolded spectrum becomes steeper at low \pt{} and
flatter at high \pt{}.  The maximum deviation of the yield for
each \pt{} bin of the unfolded spectra within the reasonable range of
$\beta$ is used as the systematic uncertainty. The uncertainty
is largest for the unbiased jet sample with resolution
parameter $R=0.3$ in the most central collisions up to 20\% at low \ptjet.

\paragraph{Unfolding method}
The spectrum obtained with the \chisq{} minimization method is compared to results using the Bayesian and SVD unfolding methods. The \chisq{} and SVD unfolded spectra agree within $\pm 10$\% for all  centrality classes and jet samples.
The Bayesian method is only included in the estimate of the
systematic uncertainties for the cases where the combinatorial jets
are suppressed by selecting jets with a leading track with $\pT > 5$
or 10 \GeVc. Without this selection, the Bayesian method was found to be
unreliable: large deviations up to $50$\% at low \ptjet{} are observed in central collisions with a resolution parameter
$R=0.3$. Such deviations are also seen in the validation studies with a heavy-ion background model
where the Bayesian method did not give the correct result, unless the truth was used as the prior. 

\paragraph{Prior}
The unfolding algorithm starts from a QCD inspired shape for the unfolded spectrum, the prior.
The measured jet spectrum is used as a standard prior for all
unfolding methods and the sensitivity to the choice of prior is
evaluated by changing the shape and yield of the prior. When the prior
is far from the truth (for example a uniform distribution), the
\chisq{} unfolding takes more iterations to converge but
eventually an unfolded jet spectrum is obtained, which is
statistically not significantly different from the unfolded spectrum
obtained with the measured spectrum as a prior. The choice of prior
has a negligible effect on the final unfolded spectrum.

\subsubsection{Combinatorial jets}
The effect of combinatorial jets in the sample is evaluated by changing
the minimum \pt{} of the unfolded spectrum and the measured range
where the unfolding is applied. 

\paragraph{Minimum \pt\ of unfolded jet spectrum}
In the default analysis the unfolded spectrum starts at $\ptjetch=0$
\GeVc. 
The sensitivity of the result to very low energy (combinatorial) jets is
explored by removing the first bin from the
unfolding procedure, i.e. starting the unfolded spectrum at $\ptjetch=5$ or
10 \GeVc{} instead of $\ptjetch=0$.
This removes one parameter
from the \chisq{} minimization. It results in an increase of the
unfolded jet yield by a few percent depending on the centrality bin
and jet radius.

\paragraph{Minimum \pt\ of measured jet spectrum}
Increasing the minimum measured \pt{} reduces the amount of combinatorial jets in the measured spectrum (see
Fig. \ref{fig:rawspectra}). The remaining combinatorial jets contribute to the jet yield at low \pt{} in the unfolded spectrum. 
The minimum \pt\ of the measured jet spectrum is varied
by $10$ \GeVc\ to a lower and higher value. With
the two variations the unfolding is performed again and the resulting
difference between the unfolded spectra with the default one assigned as a systematic uncertainty. This systematic
uncertainty is largest at low \pt\ in the region where the \ptminmeas\
cut-off is placed. For unbiased jets in most
central collisions and resolution parameter $R=0.3$ the uncertainty at
$\ptjet=40$ \GeVc\ is 25\%, while it decreases to a few percent for
$\ptjet>60$ \GeVc. 

\subsubsection{Uncertainty on background}

\paragraph{Background fluctuation distribution: random cones and high \pt\ probe embedding}
The \deltapt\ distribution obtained from embedding single high \pt\
tracks in measured \PbPb\ events is used as a variation to the
\deltapt\ distribution from random cones. The width of the background
fluctuations obtained from single-track embedding is a few 100 MeV/$c$
larger than for the random cones. The uncertainty is taken as the difference between the unfolded
jet spectrum using the \deltapt{} response from single-track embedding and the response from random cones. The difference is largest at low \ptjet{} ($<40$ \GeVc), where $\sim 15$\% deviation in the jet yield for the unbiased $R=0.3$ central jet
spectrum is observed.

\paragraph{\color{black}{Correction for collective flow effects in case of leading track biased jets}}
Due to the presence of collective effects such as elliptic and triangular flow in heavy-ion collisions the background density differs from region-to-region.
Jets with a high \pt{} leading track are preferentially found in regions with larger background density (in-plane). The subtracted background, however, is the average \pt{} density of
the event, $\rho_{\rom{ch}}$, multiplied by the area of the jet. A
correction for the larger background for biased jets
is included in the response matrix. 
This correction is determined by calculating $\rho_\rom{ch}$ on the near, away side and in the region perpendicular to the leading track biased jet in an event.
The correction is largest for events in the 10-30\% centrality class where for $R=0.3$ jets with a 5
\GeVc{} bias an overall increase of the background of 0.49 \GeVc{} is present.
The correction for flow effects is only applied for leading track biased jet spectra since for the unbiased case, jets are selected regardless of their correlation with the event or participant plane \cite{Abelev:2012ej}. 

The uncertainty on the correction for flow effects is calculated by changing the background to the lowest and highest values found in the different azimuthal regions (perpendicular and near-side regions respectively). The uncertainty on $\rho_{\rom{ch}}$ is $3$ \GeVc\ for the jet sample with a 5 \GeVc\ leading track selection, and $2$ \GeVc\ for a 10 \GeVc\ leading track requirement in central events. The systematic uncertainty on the unfolded jet spectrum for $R=0.3$ jets with $\ptleadtrack>5$ \GeVc{} in 10\% most central collisions is 8\% at $\ptjet=40$ \GeVc\ and decreases to 4\% at $\ptjet=100$ \GeVc. A previous study has shown that the background fluctuations ($\delta\pt$-distribution) are almost independent of the orientation with respect to the reaction plane \cite{Abelev:2012ej}; this effect is negligible compared to the change in the average background.

\subsubsection{Uncertainty on the detector response}
The detector response has two main components: tracking efficiency and
momentum resolution of which the tracking efficiency is the dominant uncertainty. The uncertainty on the tracking efficiency is
estimated to be $4$\%, motivated by detector simulation studies with PYTHIA and HIJING events,
and by varying the track selection criteria. To determine the systematic uncertainty on the result, a second response matrix is constructed from a simulation with a 4\% lower efficiency and the measured \PbPb{} jet spectrum is unfolded. The difference between the nominal unfolded solution
and the unfolded spectra with a modified detector response is $\sim
20$\% at $\ptjet=50$
\GeVc\ and decreases to $\sim 11$\% at $\ptjet=100$ \GeVc; the full difference is used as the systematic uncertainty, which 
corresponds to a 3--5\% uncertainty on the charged jet \pt.

\subsubsection{Centrality determination}
The relative uncertainty on the fraction of hadronic cross-section
used in the Glauber fit to determine the centrality classes is
1\% \cite{Abelev:2013qoq}. The contribution of this uncertainty on the jet spectrum is
estimated by varying the limits of the centrality classes by $\pm 1\%$
(e.g. for the 10--30\% centrality class to 9.9--29.7\% and
10.1--30.3\%). With the shifted limits of the centrality classes the
jet spectrum is compared to the nominal jet spectrum. The uncertainty
is the same for the jet spectrum with different leading track biases
and increases from central to peripheral events. For the 0--10\%
centrality class the uncertainty is less than 1\% and in the
peripheral centrality class 50--80\% it is $\sim 1.9\%$.

\begin{table}[!ht]
\begin{center}
\renewcommand{\arraystretch}{1.3}
 \begin{tabular}{>{\small}l|>{\small}l>{\small}c>{\small}c>{\small}c>{\small}c}
 & \multicolumn{1}{r}{\small{Resolution parameter}} & \multicolumn{2}{c}{\small{$R=0.2$}} & \multicolumn{2}{c}{\small{$R=0.3$}} \\
 Centrality class & \multicolumn{1}{r}{\small{\pt-interval (\GeVc)}} & 30--40 & 70--80 & 30--40 & 70--80 \\
\hline
\multirow{9}{*}{0--10\%} & Regularisation & $^{+3.4}_{-0.0}$ & $^{+2.3}_{-0.3}$ & $^{+9.9}_{-0.0}$ & $^{+2.6}_{-6.7}$\\
  & Unfolding method & $^{+0.0}_{-3.5}$ & $^{+0.0}_{-1.1}$ & $^{+0.0}_{-7.3}$ & $^{+7.6}_{-0.0}$ \\
 & Minimum \pt\ unfolded & $^{+9.6}_{-0.0}$ & $^{+0.3}_{-0.0}$ & $^{+0.0}_{-5.9}$& $^{+0.0}_{-1.8}$ \\
 & Minimum \pt\ measured & $^{+1.7}_{-4.8}$ & $^{+0.2}_{-0.3}$ & $^{+0.0}_{-13}$ & $^{+0.0}_{-2.1}$\\
 & Prior & \multicolumn{4}{c}{\small{$<0.1$}} \\
 & \deltapt & $^{+0.0}_{-4.9}$ & $^{+0.0}_{-2.1}$ & $^{+0.0}_{-27}$ & $^{+0.0}_{-4.6}$ \\
 & Detector effects & $\pm 2.7$ & $\pm 5.5$ & $\pm 4.6$ & $\pm 5.2$ \\
 & Flow bias & $^{+0.9}_{-5.8}$ & $^{+0.4}_{-4.1}$ & $^{+7.3}_{-5.9}$ & $^{+4.8}_{-4.1}$ \\
 & Centrality determination & \multicolumn{4}{c}{\small{0.8}} \\
\cline{2-6}
 & \bf{Total shape uncertainty}      & $^{+10}_{-7.6}$ & $^{+2.4}_{-2.4}$ & $^{+9.9}_{-31}$ & $^{+7.6}_{-8.6}$\\
 & \bf{Total correlated uncertainty} & $^{+2.9}_{-6.4}$ & $^{+5.6}_{-6.9}$  &  $^{+8.6}_{-7.5}$ & $^{+7.1}_{-6.6}$ \\
\hline
\hline
\multirow{9}{*}{50--80\%} & Regularisation & $^{+0.0}_{-5.5}$ & $^{+13}_{-4.1}$ & $^{+0.1}_{-5.1}$ & $^{+17}_{-2.2}$ \\
 & Unfolding method & $^{+2.1}_{-0.0}$ & $^{+0.0}_{-20}$ & $^{+2.3}_{-0.0}$ & $^{+0.0}_{-20}$ \\
 & Minimum \pt\ unfolded & $^{+0.3}_{-0.0}$ & $^{+0.1}_{-0.0}$ & $^{+1.0}_{-0.0}$ & $^{+0.6}_{-0.0}$ \\
 & Minimum \pt\ measured & $^{+9.3}_{-0.0}$ & $^{+0.7}_{-0.4}$ & $^{+7.5}_{-0.0}$ & $^{+1.0}_{-0.0}$ \\
 & Prior & \multicolumn{4}{c}{\small{$<0.1$}} \\
 & \deltapt & $^{+8.2}_{-0.0}$ & $^{+2.4}_{-0.0}$ & $^{+3.0}_{-0.0}$ & $^{+2.2}_{-0.0}$ \\ 
 & Detector effects & $\pm 3.3$ & $\pm 6.2$ & $\pm 3.3$ & $\pm 3.1$ \\
 & Flow bias & $^{+1.9}_{-1.9}$ & $^{+0.3}_{-0.3}$ & $^{+0.4}_{-7.2}$ & $^{+0.3}_{-4.0}$ \\
 & Centrality determination & \multicolumn{4}{c}{\small{1.9}} \\
\cline{2-6}
 & \bf{Total shape uncertainty}      & $^{+13}_{-5.5}$ & $^{+13}_{-20}$ & $^{+8.5}_{-5.1}$ & $^{+17}_{-20}$ \\
 & \bf{Total correlated uncertainty} & $^{+4.2}_{-4.2}$  & $^{+6.5}_{-6.5}$ & $^{+3.8}_{-8.2}$ & $^{+3.6}_{-5.4}$ \\
\hline
\end{tabular}
\end{center}
\caption{
\label{tab:SystematicsAllSources}
Overview of systematic uncertainties for jet spectra with a leading track with $\pt>5$ \GeVc. Relative uncertainties are given in percentiles for two \pt-intervals and two different centrality intervals.}
\end{table}

\subsubsection{Total systematic uncertainty}
The differential production yields are reported with their systematic uncertainties separated into two categories:
\begin{itemize}
 \item \textbf{Shape uncertainty} These are uncertainties that are
highly anti-correlated between parts of the spectrum: if the yield is
increased in some bins, it decreases in other bins. The uncertainties
from the unfolding method and regularisation, and the uncertainty on
the background fluctuations (only $\deltapt$ uncertainty) fall into
this category. The contributions are added in quadrature. 

\item \textbf{Correlated systematic uncertainty} 
These are uncertainties that result in correlated
changes over the entire spectrum. The contributions to this type of
uncertainty are the uncertainty on the detector response, the effect
of flow in the background, and the influence of the combinatorial
jets. The contributions are added in quadrature.
\end{itemize}

\subsubsection{Systematic uncertainty on ratios}
The following procedures are used for ratios of jet spectra:

\begin{itemize}
 \item \textbf{Uncorrelated uncertainties}
The systematic uncertainties from the unfolding method, which include
regularisation and variation of \pt-ranges, are not correlated from
one unfolded jet spectrum to another. The contributions from these sources are added in quadrature to calculate the uncertainies on ratios.

 \item \textbf{Correlated uncertainties}
The systematic uncertainties from the flow bias, the
\deltapt-distribution, and the detector effects are highly correlated
between unfolded spectra from different centrality classes, jet
resolution parameters and leading track biases. The uncertainty on the
tracking efficiency is similar for all centrality classes. The flow
bias depends on the \pt{} of the leading track, jet resolution
parameter, and centrality class but is correlated. As a consequence, within
a ratio the correlated systematic uncertainties
partially cancel.
\end{itemize}

\begin{figure}[tbh!f!p]
  \includegraphics[width=\textwidth]{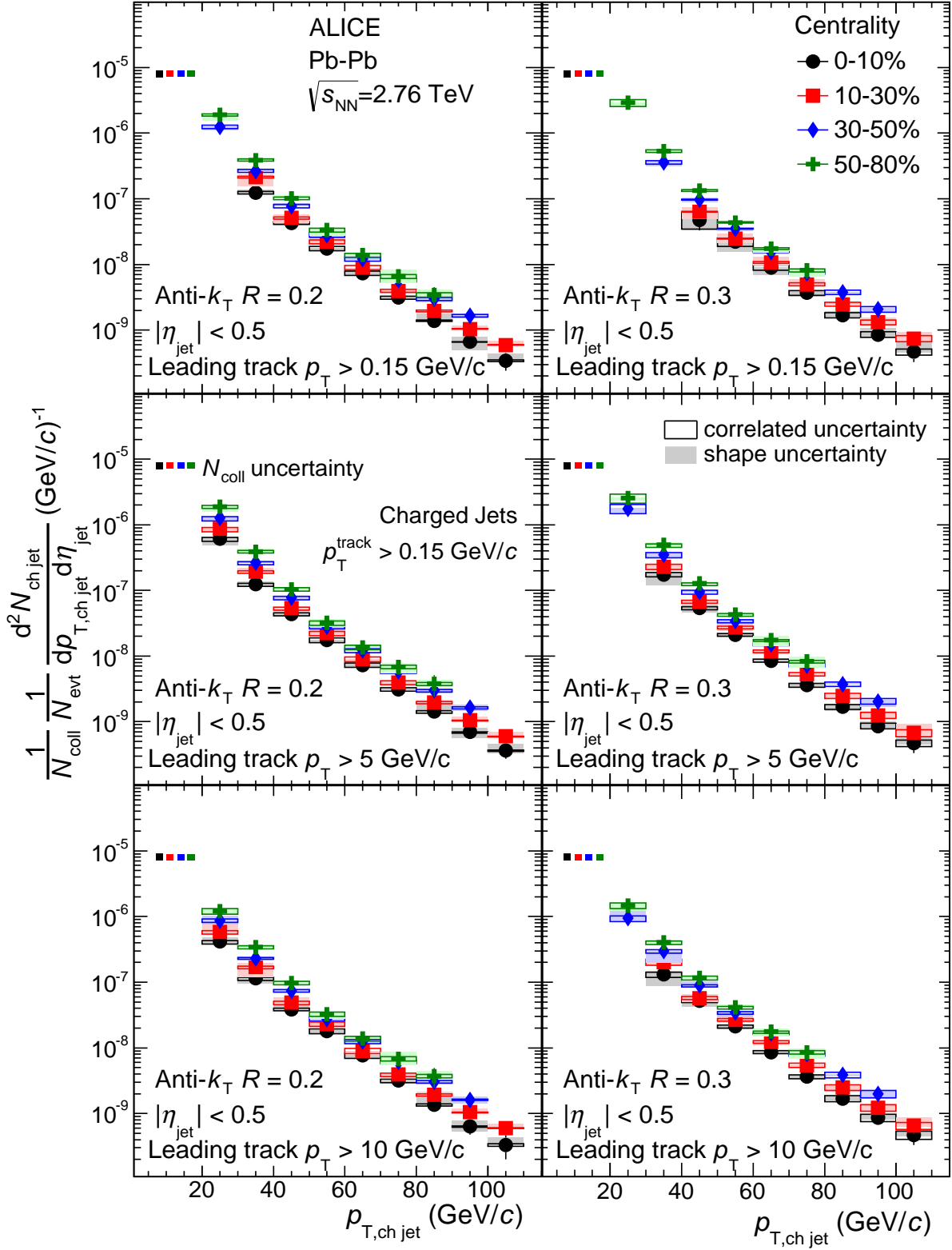}
  \caption{\label{fig:UnfoldedJetSpectraAll}Charged jet spectra, corrected for background fluctuations and detector effects, using two cone radius parameters $R=0.2$ (left panels) and $R=0.3$ (right panels) and different leading track selections: unbiased (top panels), $\ptleadtrack > 5$ \GeVc{} (middle panels), and $\ptleadtrack > 10 $ \GeVc{} (bottom panels). The uncertainty bands at $\ptjetch<20\,\GeVc$ indicate the normalisation uncertainty due to the scaling with $1/N_\mathrm{coll}$.
}
\end{figure}

\section{Results}\label{sec:results}
Jet spectra are measured with resolution parameters $R=0.2$ and $0.3$
in four centrality classes: \mbox{0--10\%}, \mbox{10--30\%}, \mbox{30--50\%} and \mbox{50--80\%}.
Figure \ref{fig:UnfoldedJetSpectraAll} shows the measured \PbPb\ jet
spectra reconstructed from charged constituents with $\pt>0.15$
\GeVc. The jet spectra are unfolded for detector effects and
background fluctuations, and corrected for the jet finding efficiency
as described in the preceding sections. The upper panels show the inclusive jet spectra while for
the center and lower panels the jet spectra with a leading track bias
of at least $5$ and $10$ \GeVc\ are shown. The markers represent the
central values of the unfolded jet spectra. It should be noted that the unfolding procedure leads to correlations between the data points, because the width of the response function is similar to the bin width: neighboring \pt-bins tend to
fluctuate together (correlated) while bins with some distance tend to
be anti-correlated. The vertical error bars represent the statistical
uncertainties. The filled and
open boxes indicate the corresponding shape and correlated systematic uncertainties discussed previously.

The jet yield is given per event and normalized by the average number
of nucleon-nucleon collisions $N_{\rom{coll}}$ corresponding to the
given centrality interval. The markers shown below 20 \GeVc\ indicate
the normalization uncertainty on the extracted values of
$N_{\rom{coll}}$ (see Table \ref{tab:NCollTAANPart}). The jet yield
evolves with centrality: for central collisions fewer jets are
observed per $N_{\rom{coll}}$ than in peripheral collisions.

\begin{figure}[tbh!f!p]
\centering
\includegraphics[width=\textwidth]{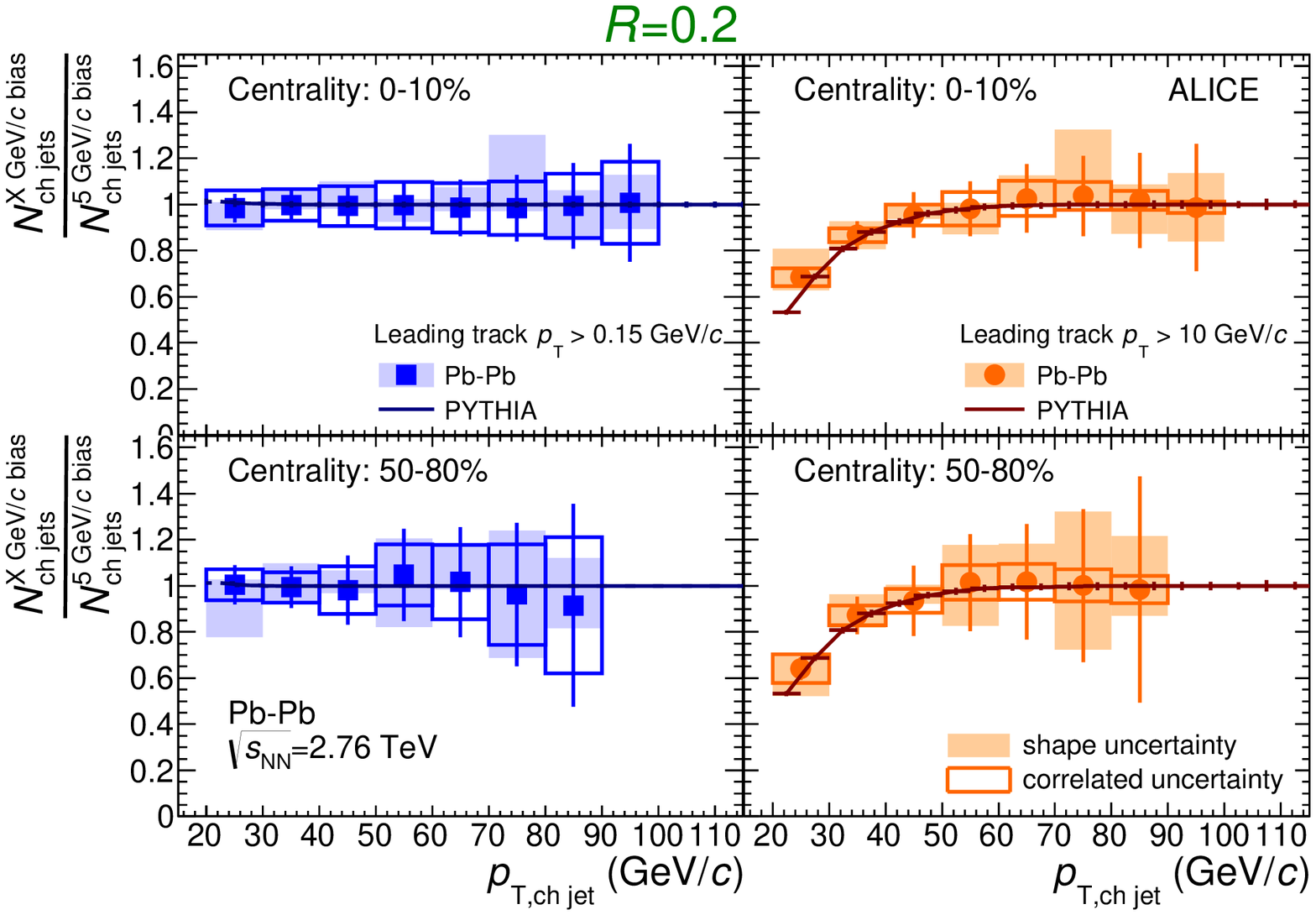}
\includegraphics[width=\textwidth]{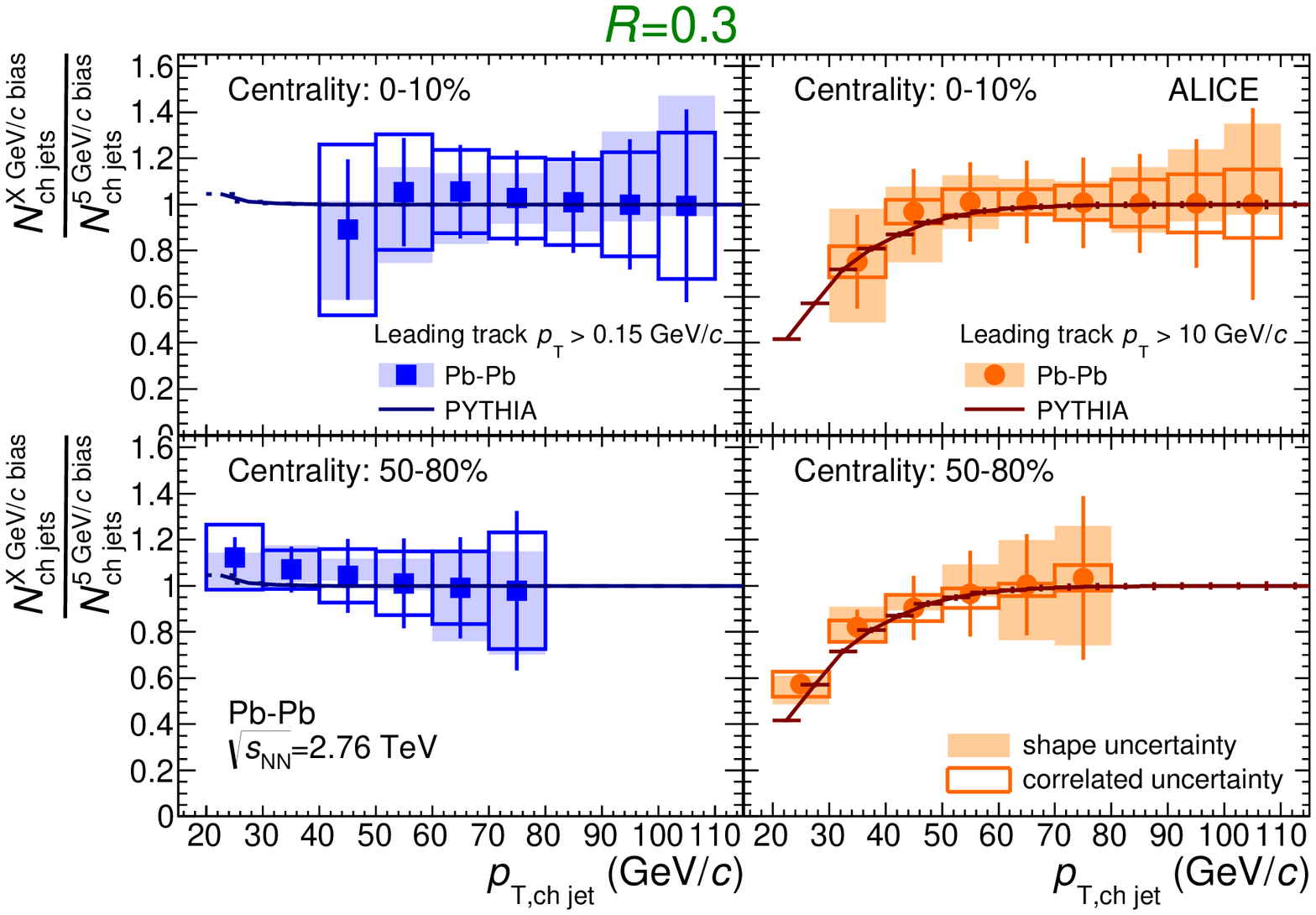}
\caption{\label{fig:biased_unbiased} Ratio of reconstructed unbiased and leading track biased jet yields for two resolution parameters (top panel: $R=0.2$; bottom panel: $R=0.3$). Calculations of the same ratio with the PYTHIA model (particle level) are shown for reference. Left panels: ratio of unbiased spectra to $\ptleadtrack > 5$ \GeVc. Right panels: ratio of spectra with $\ptleadtrack > 10$ \GeVc{}  to $\ptleadtrack > 5$ \GeVc.
}
\end{figure}

The left panels in Fig. \ref{fig:biased_unbiased} show the ratio
between the unbiased jet spectra and jets with a leading track of at
least $5$ \GeVc. Although the biased spectrum is a subset of the unbiased spectrum, the statistical uncertainties are added in quadrature since the unfolding procedure introduces a point-to-point correlation between the statistical uncertainties. In the jet \pt-range considered here, $\ptjetch>20$
\GeVc, the PYTHIA vacuum expectation from the Perugia-2011 tune \cite{Skands:2010ak} is that almost all jets have a
constituent of at least $5$ \GeVc, resulting in a ratio at unity as
indicated by the PYTHIA data points. The ratio between the
unbiased and $5$ \GeVc\ biased measured \PbPb\ jet spectra is
consistent in peripheral and central collisions with the vacuum
expectation. No evidence of the modification
of the hard jet core is observed.

The right panels in Fig. \ref{fig:biased_unbiased} show the ratio
between the jet spectra with a leading track \pt\ of at least $10$
\GeVc\ and $5$ \GeVc\ as measured in central and peripheral \PbPb\
collisions compared to the same observable at particle level in
PYTHIA with the Perugia-2011 tune. By selecting jets with a
higher momentum for the leading jet constituent, low \pt\ jets with a
soft fragmentation pattern are removed from the sample. The ratio
increases with \pt\ reaching unity at $\ptjetch=50$ \GeVc\ for $R=0.2$
jets and at $\ptjetch=60$ \GeVc\ for $R=0.3$ jets in central and
peripheral collisions. This rising trend is due to the increased fragmentation bias and is compatible with the
fragmentation bias observed in PYTHIA. 

\begin{figure}[tbh!f]
  \includegraphics[width=0.5\textwidth]{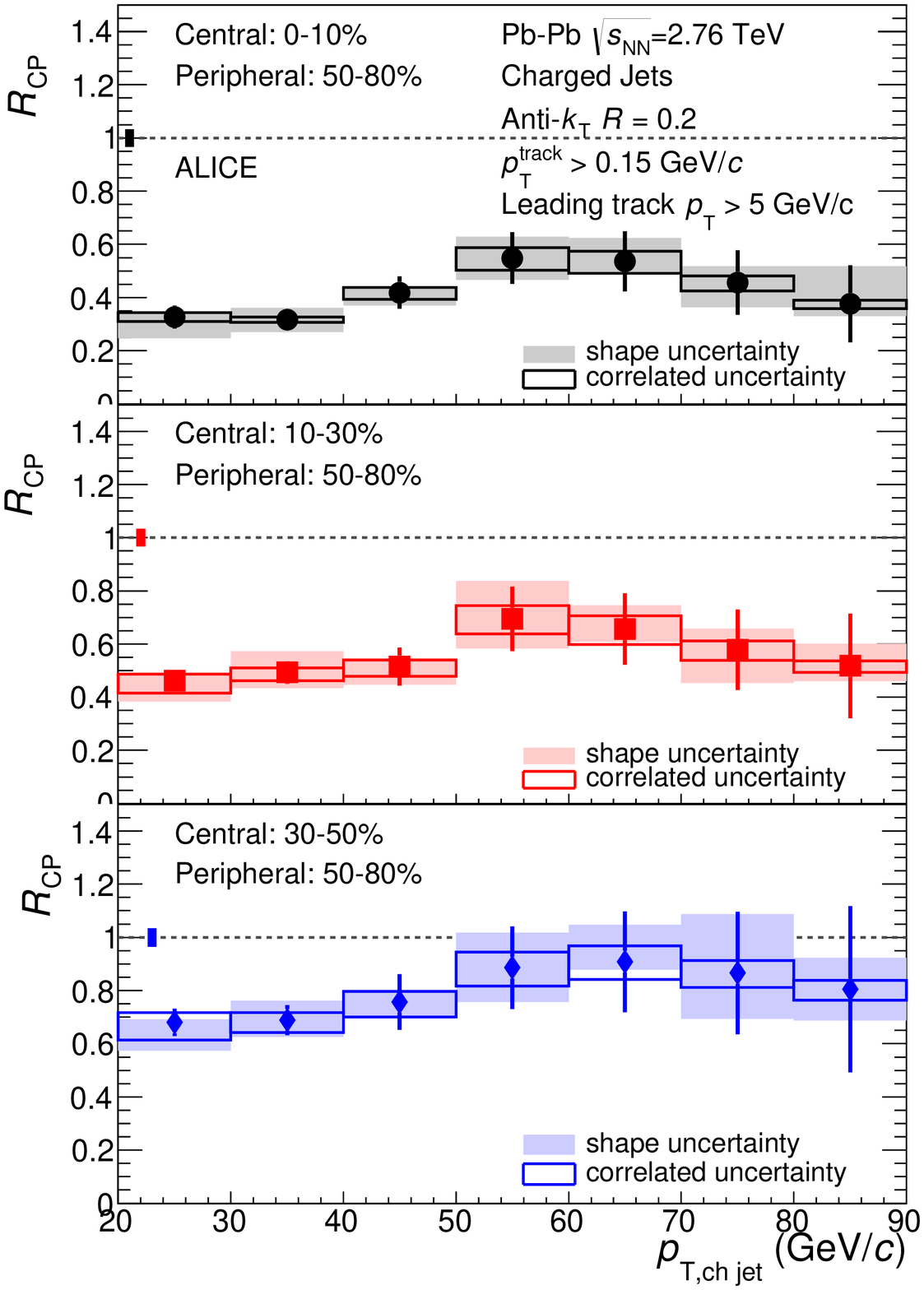}
  \includegraphics[width=0.5\textwidth]{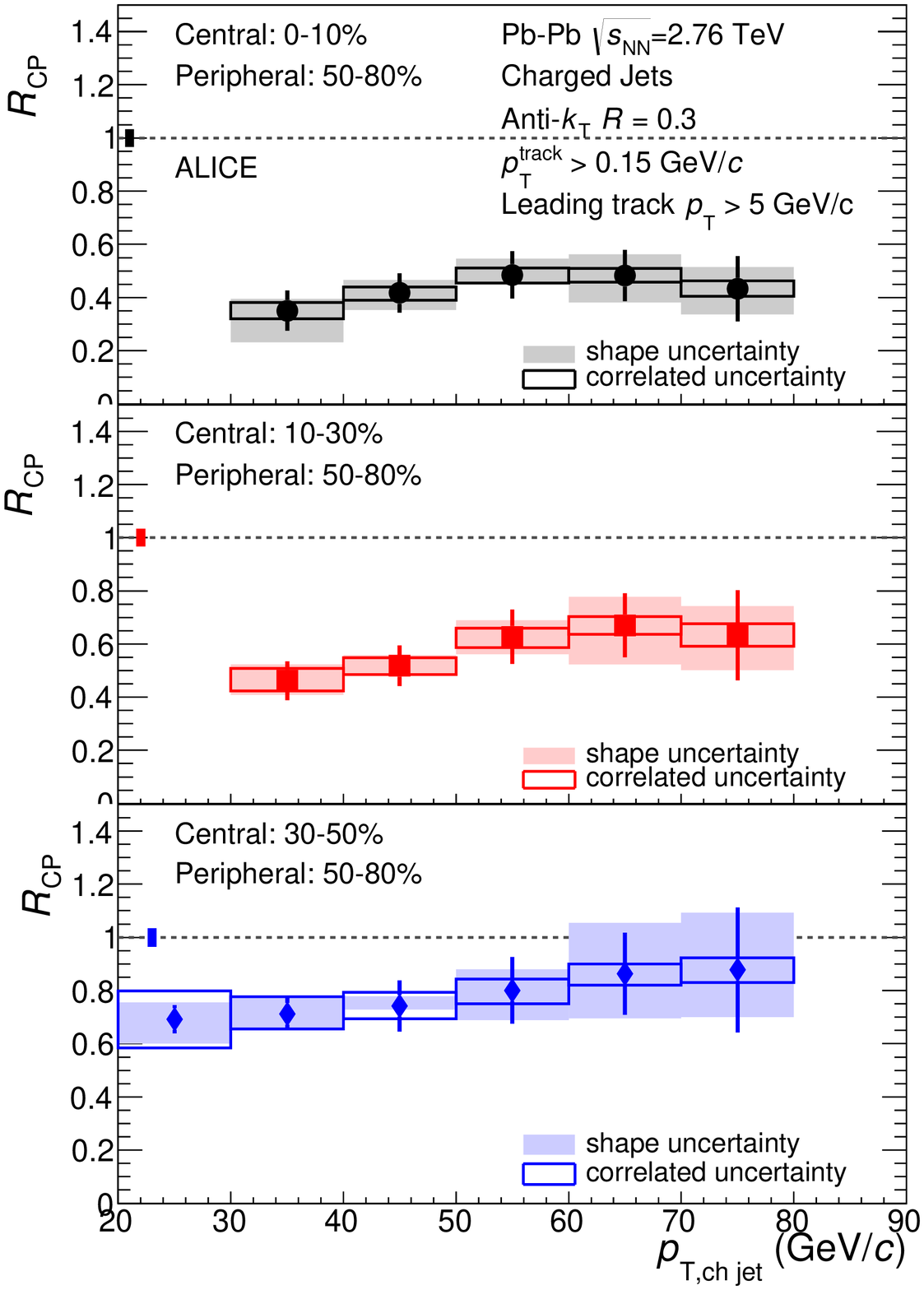}
\caption{\label{fig:RCP}Nuclear modification factor \RCP{} for charged jets with a leading charged particle with $\pTleading > 5$ \GeVc,  with $R=0.2$ (left panels) and $R=0.3$ (right panels) and different centrality selections.} 
\end{figure}

Jet quenching in most central heavy-ion collisions is quantified by constructing the jet nuclear modification factor \RCP{}, 
\begin{equation}\label{eq:RCP}
{\Large\RCP} = \dfrac{\dfrac{1}{\langle T_{\rom{AA}}\rangle}\dfrac{1}{N_{\rom{evt}}}\dfrac{\rom{d}^{2}N_{\rom{ch\;jet}}}{\rom{d}\ptjetch\rom{d}\eta_{\rom{ch\;jet}}}\bigg|_\rom{central}}{\dfrac{1}{\langle T_{\rom{AA}}\rangle}\dfrac{1}{N_{\rom{evt}}}\dfrac{\rom{d}^{2}N_{\rom{ch\;jet}}}{\rom{d}\ptjetch\rom{d}\eta_{\rom{ch\;jet}}}\bigg|_{\rom{peripheral}}},
\end{equation}
which is the ratio of jet \pt\ spectra in central and peripheral collisions normalized by the nuclear overlap functions
$\langle T_{\rom{AA}}\rangle$ as
calculated with a Glauber model for each centrality class\cite{Abelev:2013qoq}. If the full
jet energy is recovered within the cone, and in the
absence of initial state effects like parton shadowing \cite{Amaudruz:1995tq,Eskola:2009uj,Cronin:1974zm}, $\RCP$ is unity by
construction. In that case, jet quenching would manifest itself as
redistribution of the energy within the cone as compared to jet
fragmentation in the vacuum. The jet
suppression factor \RCP\ is shown in Fig. \ref{fig:RCP}, using 
centrality class 50-80\%  as the peripheral reference. A strong jet suppression, $0.3< \RCP < 0.5$, is observed
for 0-10\% central events, while more
peripheral collisions (30-50\%) are less suppressed, $\RCP\simeq0.8$ at high \ptjetch. A mild
increase of \RCP\ with increasing \ptjetch{} is observed at low jet energies
while at high $\pt\gtrsim 50$ \GeVc{} the suppression is consistent with a constant. The \RCP{} does not change significantly with the resolution parameter $R$ for the range studied ($R=0.2$ and $R=0.3$).
\begin{figure}[tbh!f]
  \includegraphics[width=0.5\textwidth]{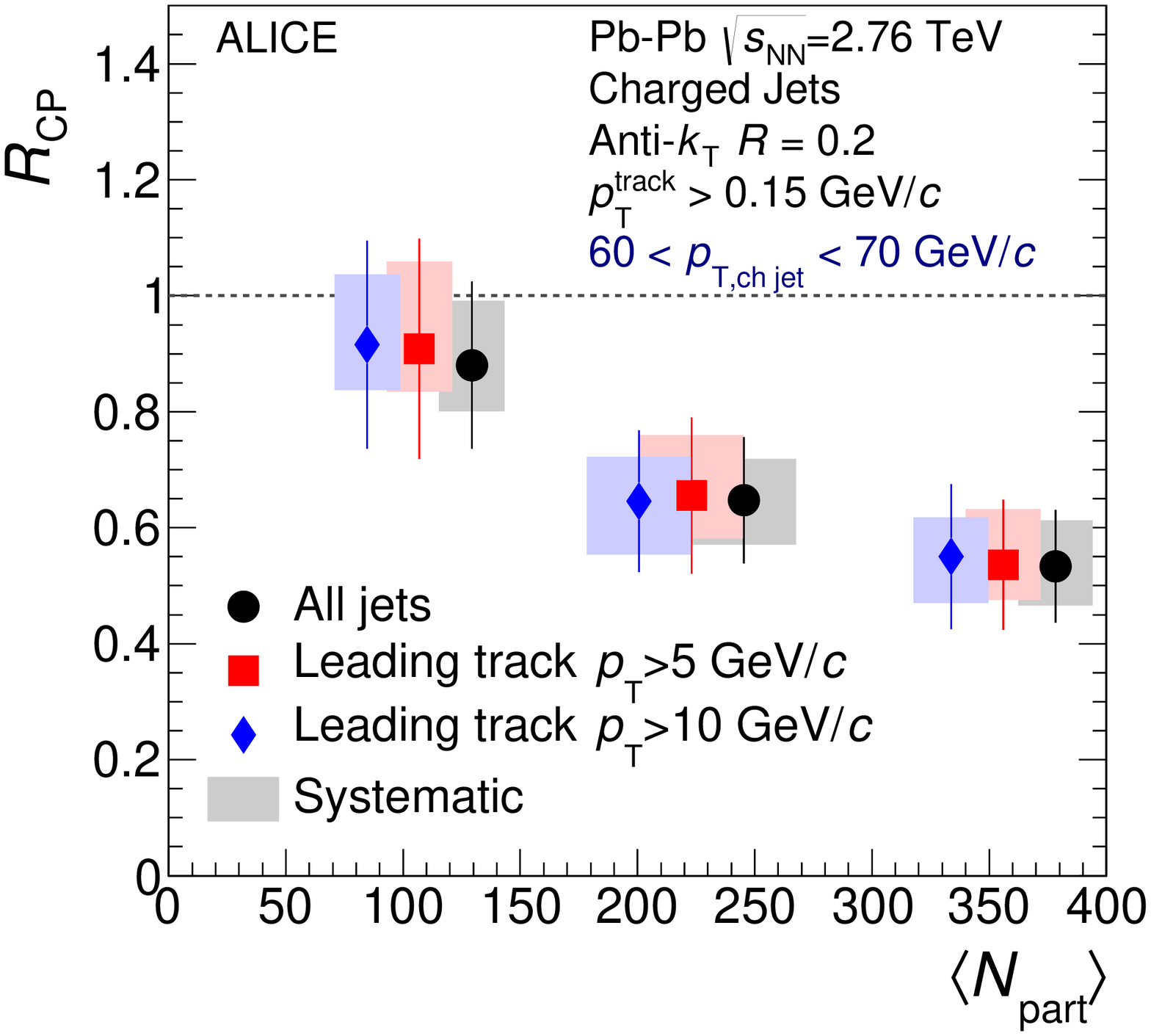}
  \includegraphics[width=0.5\textwidth]{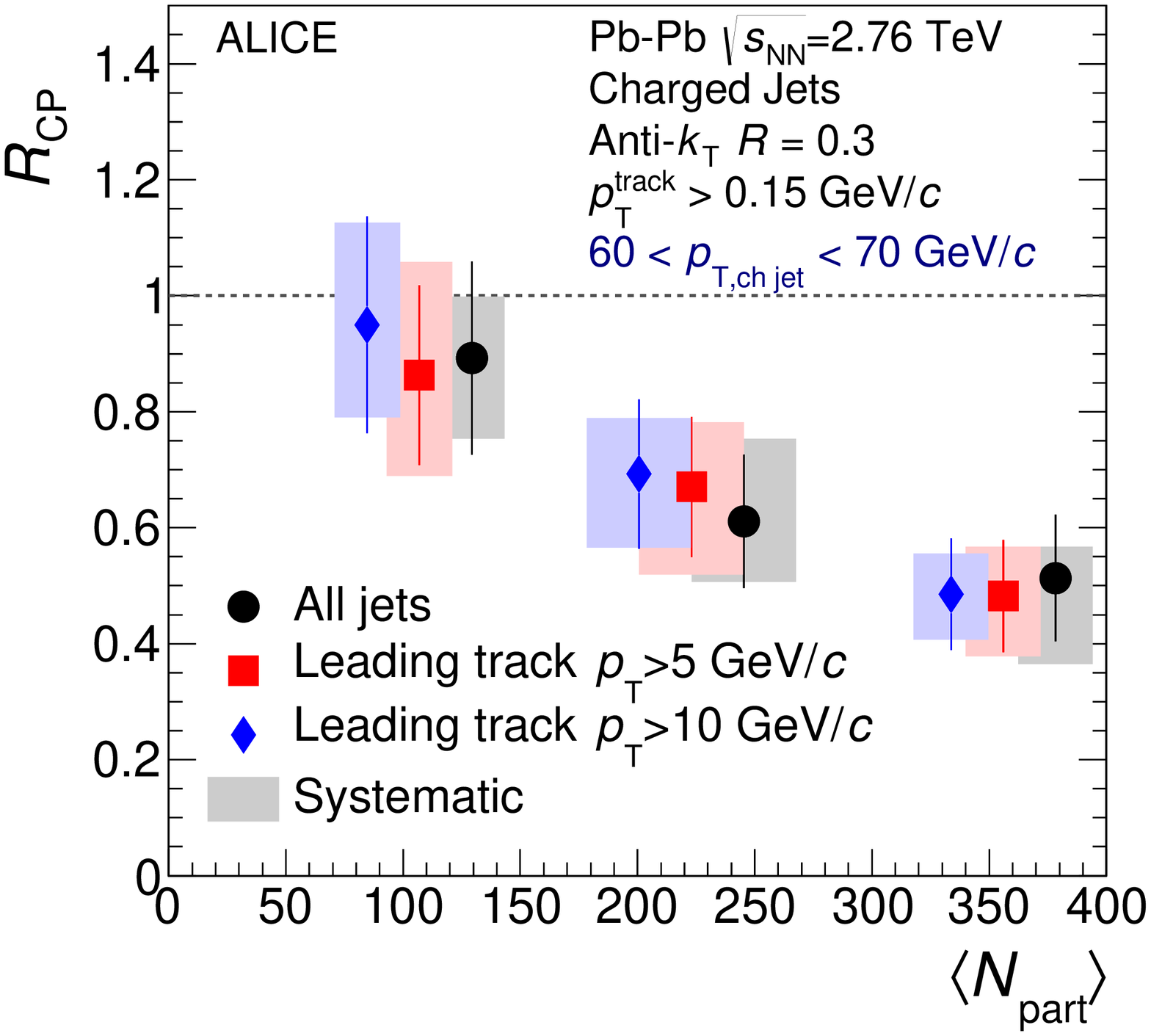}
\caption{\label{fig:RCPNPart}
\RCP{} for unbiased and leading track biased jets with $60<\ptjetch<70$ \GeVc{} as a function of the average number of participants in the collision.  Left panel: $R=0.2$. Right panel: $R=0.3$. For visibility the data points for all jets and for jets with $\pTleading > 10$ \GeVc{} are shifted to the left and right respectively.
}
\end{figure}

Figure \ref{fig:RCPNPart} shows the jet \RCP\ at $60<\ptjetch<70$ \GeVc\ as a function of the average number of participant nucleons corresponding to the selected centrality classes (see Table \ref{tab:NCollTAANPart}). 
A decreasing trend of the \RCP\ as a function of the number of participants is observed.
Figure \ref{fig:RCPNPart} also compares the suppression of jets with a high \pt{} track selection, and shows  no evident dependence on the fragmentation pattern.

\begin{figure}[tbh!f]
 \centering
  \includegraphics[width=1.3\myfigwidth]{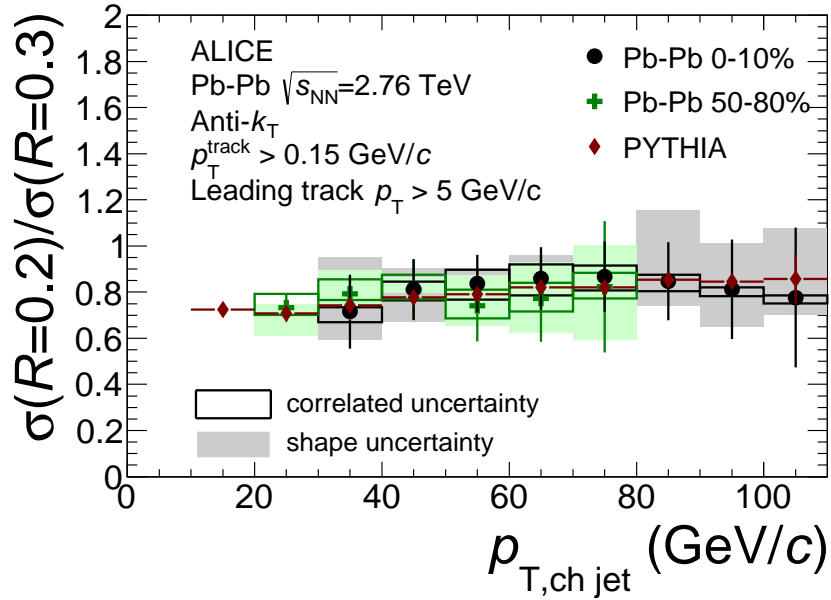}
\caption{\label{fig:RatioR02R03CompPYTHIA}Ratio of charged jet
  \pT-spectra with radius parameter $R=0.2$ and $0.3$ and a leading
  charged particle $\pTleading > 5$ \GeVc\ in \PbPb{} data and simulated
  PYTHIA events.}
\end{figure}

The ratio of the jet \pt\ spectra measured at different $R$ can
potentially provide information about jet structure modifications due
to redistribution of energy caused by jet quenching
\cite{Vitev:2008bx,Vitev:2008rz}. Figure
\ref{fig:RatioR02R03CompPYTHIA} shows the measured ratio
$\sigma(R=0.2)/\sigma(R=0.3)$ for central and peripheral
collisions.  The comparison of the measured ratio to the ratio obtained with PYTHIA (particle level) shows that the transverse jet shape in central and peripheral \PbPb{} collisions are consistent with jet shapes in vacuum. 
No sign of a modified jet
structure is observed between radii of 0.2 and 0.3 within uncertainties.

\section{Discussion and Conclusions}\label{sec:discussion}

Before the first jet measurements in heavy ion collisions were performed, it was expected that medium interactions redistribute the momenta of jet fragments to small or moderate angles, because of kinematic effects (the momentum of the jet is large compared to the typical
momenta of partons in the medium) as well as dynamics (the cross
section for medium-induced radiation peaks at small angles \cite{Salgado:2003gb}). 
At the same time, there were some indications from numerical calculations by Vitev \cite{Vitev:2005yg} and in the q-PYTHIA event generator 
\cite{Armesto:2009fj,Apolinario:2012cg} that large angle radiation is kinematically favoured for large medium density.
The first jet measurements in heavy ion collisions at the LHC showed a large energy imbalance for jet pairs \cite{Aad:2010bu,Chatrchyan:2011sx,Chatrchyan:2012nia}, indicating that a significant fraction of jet momentum is transported out of the jet cone by interactions with the medium for recoil jets. Since then, it has been realised that there is a variety of mechanisms that may contribute to large angle
radiation, such as jet broadening by medium-induced virtuality (YaJEM)
\cite{Renk:2009nz,Renk:2010zx}, reinteractions of the radiated gluons (also
called 'frequency collimation of the radiation')
\cite{CasalderreySolana:2010eh,Renk:2012cx}, and quantum
(de-)coherence effects \cite{MehtarTani:2012cy,MehtarTani:2011tz}.

The large suppression of charged jet production with $R=0.2$ and $R=0.3$ in central \PbPb{}
collisions shown in Fig.~\ref{fig:RCP}, also indicates that momentum transport to large angles is an important effect.

\begin{figure}
\centering
\includegraphics[width=1.3\myfigwidth]{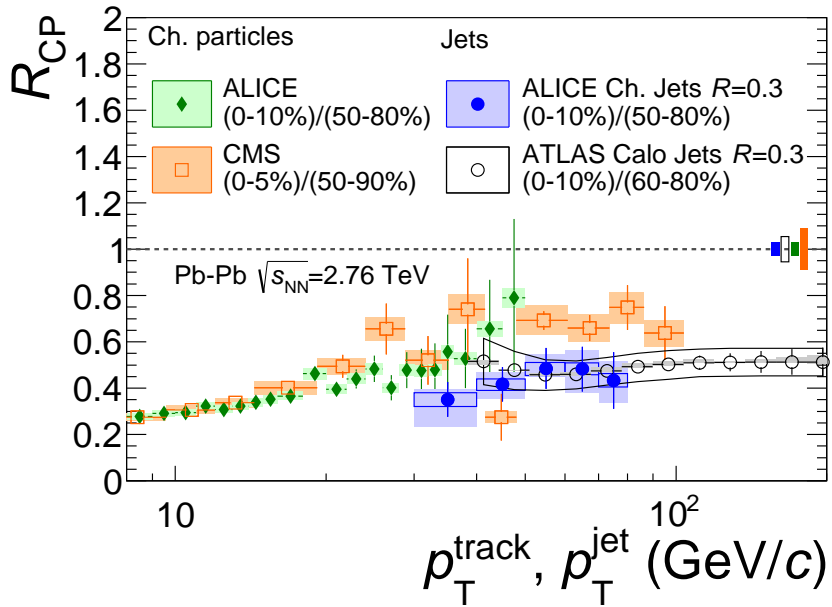}
\caption{\label{fig:compareAll}Comparison to jet \RCP{} measured by ATLAS \cite{Aad:2012vca} and to charged particle suppression by ALICE \cite{Abelev:2012eq} and CMS \cite{CMS:2012aa}. Note that the underlying parton $\pt$ scale is different at the same measured $\pt$ for charged particles, charged jets and full jets.}
\end{figure}

To further explore these effects, Fig.~\ref{fig:compareAll} compares
the jet measurement reported in this paper to the nuclear modification factor for charged
hadrons measured by ALICE \cite{Abelev:2012eq} and CMS
\cite{CMS:2012aa} and to the calorimetric jet measurements by ATLAS
\cite{Aad:2012vca}.
 
Comparing the \RCP{} of jets to charged particles in Fig.~\ref{fig:compareAll}, one would expect the suppression for jets
to be smaller than for hadrons, since jet reconstruction collects
multiple jet fragments into the jet cone, thus recovering some of the
medium-induced fragmentation. However, it can be seen that the \RCP{}
for jets is similar to that observed for single hadrons over a broad momentum range. This
indicates that the momentum is redistributed to angles larger than $R=0.3$ by interactions with the
medium.

Such a strong redistribution of momentum might also be expected to lead
to a significant broadening of the energy profile within the larger
cone radius $R=0.3$. The results presented in this paper,
however,
show that the ratio of yields for jets with $R=0.2$ and $R=0.3$ is
similar in PYTHIA \pp{} simulations and \PbPb{} collisions (see 
Fig.~\ref{fig:RatioR02R03CompPYTHIA}), indicating that the energy
profile of the found jets is not significantly modified. In addition, Fig.~\ref{fig:biased_unbiased} shows that the
effect of selecting jets with a leading hadron with $\pT > 5$ or
$10\;\GeVc$ is similar in \PbPb{} collisions and in PYTHIA \pp{}
events, 
which indicates that the longitudinal momentum distribution of (leading) high \pt{} tracks in jets reconstructed in Pb-Pb collisions remains largely unmodified. This observation is in qualitative agreement with measurements of fragmentation properties by CMS \cite{Chatrchyan:2012gw,Chatrchyan:2013kwa} and ATLAS \cite{ATLAS_ff_conf}.

A further impression of the importance of soft radiation can be obtained
by comparing the calorimetric jet measurement by ATLAS to the
ALICE results in this paper. The ALICE measurement is more sensitive to low-momentum
fragments due to the high tracking efficiency and good momentum
resolution of charged particle tracks at low \pt. The agreement
between these two jet measurements in Fig.~\ref{fig:compareAll}
suggests that the contribution of low momentum fragments to the jet energy is small. A study of PYTHIA events shows that the
expected contribution of fragments with $\pT < 1(2)$ \GeVc{} is 4(7)\%
of the jet energy at $\ptjetch=40$ \GeVc{} with cone radius
$R=0.2(0.3)$ in \pp{} collisions. The results indicate that this
contribution is also limited in \PbPb{} collisions.

The measured ratios of jet cross sections with $R=0.2$ and
$R=0.3$ and with and without leading particle selection show that the
transverse and longitudinal fragment distributions of the
reconstructed jets are similar in \pp{} (PYTHIA calculations) and
\PbPb{} collisions. This 'unmodifed hard core' of the jet may be due to
formation time effects (the parton leaves the medium with relatively
high momentum and then fragments without further interactions)
\cite{Renk:2009nz,CasalderreySolana:2011gx}, quantum interference
effects (a group of partons with small opening angles interacts with
the medium as one parton)\cite{Mehtar-Tani:2013pia}, kinematics (large
momentum emissions are kinematically favoured at small angles)
\cite{Renk:2009nz} and/or selection bias effects
\cite{Renk:2009nz,Renk:2012ve}.

First results from the JEWEL event generator show a strong suppression
of jets, in agreement with the \RCP{} shown in
Fig.~\ref{fig:compareAll} \cite{Zapp:2012ak}. However, more extensive
comparisons of theoretical models to the different experimental measurements are needed to
determine how well they constrain the dynamics of parton energy loss
models.

In summary, a measurement of charged jet spectra in
\PbPb{} collisions at different centralities was reported, using charged hadrons
with $\pT > 0.15$ GeV/$c$.  The analysis was performed for a jet sample
with a minimal fragmentation bias by introducing different \pt-ranges in
the unfolding procedure for the unfolded and measured spectrum.  To
suppress combinatorial jets from the measured population, jet spectra with a
leading track selection of $\ptleadtrack>5$ and 10 \GeVc{} were also
reported.  The effect of the leading track cut at 5
\GeVc{} is small for the measured range \mbox{$\ptjetch > 20$ \GeVc},
while for \mbox{$\ptleadtrack>10$ \GeVc}, the effect is sizeable, but
consistent with expectations from jet
fragmentation in PYTHIA events, indicating that the high-\pT{}
fragmentation is not strongly modified by interactions with the
medium. The ratio of jets reconstructed with $R=0.2$ and $R=0.3$ is
found to be similar in central and peripheral \PbPb\ events, and
similar to PYTHIA calculations, indicating no strong broadening of the
radial jet profile within $R=0.3$. The nuclear modification factor \RCP\ for jets is
in the range 0.3--0.5, and tends to be lower at low $\ptjetch \approx
30$ \GeVc\ than at high $\ptjetch \approx 100$ \GeVc. The value of
\RCP\ for jets is similar to charged hadrons, which suggests that
interactions with the medium redistribute energy and momentum to
relatively large angles with respect to the jet axis.

\newenvironment{acknowledgement}{\relax}{\relax}
\begin{acknowledgement}
\section*{Acknowledgements}
The ALICE collaboration would like to thank all its engineers and technicians for their invaluable contributions to the construction of the experiment and the CERN accelerator teams for the outstanding performance of the LHC complex.
The ALICE collaboration gratefully acknowledges the resources and support provided by all Grid centres and the Worldwide LHC Computing Grid (WLCG) collaboration.
The ALICE collaboration acknowledges the following funding agencies for their support in building and
running the ALICE detector:
State Committee of Science,  World Federation of Scientists (WFS)
and Swiss Fonds Kidagan, Armenia,
Conselho Nacional de Desenvolvimento Cient\'{\i}fico e Tecnol\'{o}gico (CNPq), Financiadora de Estudos e Projetos (FINEP),
Funda\c{c}\~{a}o de Amparo \`{a} Pesquisa do Estado de S\~{a}o Paulo (FAPESP);
National Natural Science Foundation of China (NSFC), the Chinese Ministry of Education (CMOE)
and the Ministry of Science and Technology of China (MSTC);
Ministry of Education and Youth of the Czech Republic;
Danish Natural Science Research Council, the Carlsberg Foundation and the Danish National Research Foundation;
The European Research Council under the European Community's Seventh Framework Programme;
Helsinki Institute of Physics and the Academy of Finland;
French CNRS-IN2P3, the `Region Pays de Loire', `Region Alsace', `Region Auvergne' and CEA, France;
German BMBF and the Helmholtz Association;
General Secretariat for Research and Technology, Ministry of
Development, Greece;
Hungarian OTKA and National Office for Research and Technology (NKTH);
Department of Atomic Energy and Department of Science and Technology of the Government of India;
Istituto Nazionale di Fisica Nucleare (INFN) and Centro Fermi -
Museo Storico della Fisica e Centro Studi e Ricerche "Enrico
Fermi", Italy;
MEXT Grant-in-Aid for Specially Promoted Research, Ja\-pan;
Joint Institute for Nuclear Research, Dubna;
National Research Foundation of Korea (NRF);
CONACYT, DGAPA, M\'{e}xico, ALFA-EC and the EPLANET Program
(European Particle Physics Latin American Network)
Stichting voor Fundamenteel Onderzoek der Materie (FOM) and the Nederlandse Organisatie voor Wetenschappelijk Onderzoek (NWO), Netherlands;
Research Council of Norway (NFR);
Polish Ministry of Science and Higher Education;
National Science Centre, Poland;
 Ministry of National Education/Institute for Atomic Physics and CNCS-UEFISCDI - Romania;
Ministry of Education and Science of Russian Federation, Russian
Academy of Sciences, Russian Federal Agency of Atomic Energy,
Russian Federal Agency for Science and Innovations and The Russian
Foundation for Basic Research;
Ministry of Education of Slovakia;
Department of Science and Technology, South Africa;
CIEMAT, EELA, Ministerio de Econom\'{i}a y Competitividad (MINECO) of Spain, Xunta de Galicia (Conseller\'{\i}a de Educaci\'{o}n),
CEA\-DEN, Cubaenerg\'{\i}a, Cuba, and IAEA (International Atomic Energy Agency);
Swedish Research Council (VR) and Knut $\&$ Alice Wallenberg
Foundation (KAW);
Ukraine Ministry of Education and Science;
United Kingdom Science and Technology Facilities Council (STFC);
The United States Department of Energy, the United States National
Science Foundation, the State of Texas, and the State of Ohio.
\end{acknowledgement}

\enlargethispage{0.5cm}
\bibliographystyle{utphys}
\bibliography{biblionodoi}

\providecommand{\href}[2]{#2}\begingroup\raggedright\begin{thebibliography}{10}

\bibitem{Karsch:2003jg}
F.~Karsch and E.~Laermann, ``{Thermodynamics and in-medium hadron properties
  from lattice QCD},'' in {\em Quark-Gluon Plasma 3}, R.~C. Hwa, ed.,
  pp.~1--59.
\newblock World Scientific, 2003.
\newblock
\href{http://arxiv.org/abs/hep-lat/0305025}{{\ttfamily arXiv:hep-lat/0305025}}.
\newblock

\bibitem{Bazavov:2011nk}
A.~Bazavov {\em et~al.}, ``{The chiral and deconfinement aspects of the QCD
  transition},'' {\em Phys. Rev.} {\bfseries D85} (2012) 054503,
\href{http://arxiv.org/abs/1111.1710}{{\ttfamily arXiv:1111.1710 [hep-lat]}}.

\bibitem{Gyulassy:1990ye}
M.~Gyulassy and M.~Plumer, ``{Jet quenching in dense matter},''
\href{http://dx.doi.org/10.1016/0370-2693(90)91409-5}{{\em Phys.Lett.}
  {\bfseries B243} (1990) 432--438}.

\bibitem{Baier:1994bd}
R.~Baier, Y.~L. Dokshitzer, S.~Peigne, and D.~Schiff, ``{{Induced gluon
  radiation in a QCD medium}},''
  \href{http://dx.doi.org/10.1016/0370-2693(94)01617-L}{{\em Phys.Lett.}
  {\bfseries B345} (1995) 277--286},
\href{http://arxiv.org/abs/hep-ph/9411409}{{\ttfamily arXiv:hep-ph/9411409
  [hep-ph]}}.

\bibitem{Adcox:2001jp}
{\bfseries PHENIX} Collaboration, K.~Adcox {\em et~al.}, ``{Suppression of
  hadrons with large transverse momentum in central Au+Au collisions at
  $\sqrt{s_{NN}}$ = 130-GeV},''
  \href{http://dx.doi.org/10.1103/PhysRevLett.88.022301}{{\em Phys.Rev.Lett.}
  {\bfseries 88} (2002) 022301},
\href{http://arxiv.org/abs/nucl-ex/0109003}{{\ttfamily arXiv:nucl-ex/0109003
  [nucl-ex]}}.

\bibitem{Adler:2003qi}
{\bfseries PHENIX} Collaboration, S.~S. Adler {\em et~al.}, ``{Suppressed
  $\pi^0$ production at large transverse momentum in central Au + Au collisions
  at $\sqrt{s_{NN}}$ = 200 GeV},''
  \href{http://dx.doi.org/10.1103/PhysRevLett.91.072301}{{\em Phys. Rev. Lett.}
  {\bfseries 91} (2003) 072301},
\href{http://arxiv.org/abs/nucl-ex/0304022}{{\ttfamily arXiv:nucl-ex/0304022}}.

\bibitem{Ada03b}
{\bfseries STAR} Collaboration, J.~Adams {\em et~al.}, ``{Transverse Momentum
  and Collision Energy Dependence of High $p_T$ Hadron Suppression in Au+Au
  Collisions at Ultrarelativistic Energies},'' {\em Phys. Rev. Lett.}
  {\bfseries 91} (2003) 172302,
\href{http://arxiv.org/abs/nucl-ex/0305015}{{\ttfamily nucl-ex/0305015}}.

\bibitem{Adams:2003im}
{\bfseries STAR} Collaboration, J.~e.~a. Adams, ``Evidence from
  $d+\mathrm{A}\mathrm{u}$ measurements for final-state suppression of
  high-${p}_{T}$ hadrons in $\mathrm{A}\mathrm{u}+\mathrm{A}\mathrm{u}$
  collisions at rhic,''
  \href{http://dx.doi.org/10.1103/PhysRevLett.91.072304}{{\em Phys. Rev. Lett.}
  {\bfseries 91} (Aug, 2003) 072304}.
  \url{http://link.aps.org/doi/10.1103/PhysRevLett.91.072304}.

\bibitem{Arsene:2003yk}
{\bfseries BRAHMS} Collaboration, I.~Arsene {\em et~al.}, ``{Transverse
  momentum spectra in Au+Au and d+Au collisions at s**(1/2) = 200-GeV and the
  pseudorapidity dependence of high p(T) suppression},''
  \href{http://dx.doi.org/10.1103/PhysRevLett.91.072305}{{\em Phys.Rev.Lett.}
  {\bfseries 91} (2003) 072305},
\href{http://arxiv.org/abs/nucl-ex/0307003}{{\ttfamily arXiv:nucl-ex/0307003
  [nucl-ex]}}.

\bibitem{Back:2003qr}
{\bfseries PHOBOS} Collaboration, B.~Back {\em et~al.}, ``{Charged hadron
  transverse momentum distributions in Au + Au collisions at (S(NN))**(1/2) =
  200-GeV},'' \href{http://dx.doi.org/10.1016/j.physletb.2003.10.101}{{\em
  Phys.Lett.} {\bfseries B578} (2004) 297--303},
\href{http://arxiv.org/abs/nucl-ex/0302015}{{\ttfamily arXiv:nucl-ex/0302015
  [nucl-ex]}}.

\bibitem{Aamodt:2010jd}
{\bfseries ALICE} Collaboration, K.~Aamodt {\em et~al.}, ``{Suppression of
  charged particle production at large transverse momentum in central Pb--Pb
  collisions at $\sqrt{s_{NN}}$ = 2.76 TeV },''
  \href{http://dx.doi.org/http://dx.doi.org/10.1016/j.physletb.2010.12.020}{{\em
  Physics Letters B} {\bfseries 696} no.~12, (2011) 30 -- 39}.

\bibitem{Aamodt:2011vg}
{\bfseries ALICE} Collaboration, K.~Aamodt {\em et~al.}, ``{Particle-yield
  modification in jet-like azimuthal di-hadron correlations in Pb-Pb collisions
  at $\sqrt{s_{NN}} = 2.76$ TeV},''
  \href{http://dx.doi.org/10.1103/PhysRevLett.108.092301}{{\em Phys.Rev.Lett.}
  {\bfseries 108} (2012) 092301},
\href{http://arxiv.org/abs/1110.0121}{{\ttfamily arXiv:1110.0121 [nucl-ex]}}.

\bibitem{ATLAS-CONF-2012-120}
{\bfseries ATLAS} Collaboration, ``{Measurement of the charged particle spectra
  in PbPb collisions at sqn=2.76TeV with the ATLAS detector at the LHC},''
  Tech. Rep. ATLAS-CONF-2012-120, CERN, Geneva, Aug, 2012.

\bibitem{CMS:2012aa}
{\bfseries CMS} Collaboration, S.~Chatrchyan {\em et~al.}, ``{Study of high-pT
  charged particle suppression in PbPb compared to $pp$ collisions at
  $\sqrt{s_{NN}}=2.76$ TeV},''
  \href{http://dx.doi.org/10.1140/epjc/s10052-012-1945-x}{{\em Eur.Phys.J.}
  {\bfseries C72} (2012) 1945},
\href{http://arxiv.org/abs/1202.2554}{{\ttfamily arXiv:1202.2554 [nucl-ex]}}.

\bibitem{Aad:2010bu}
{\bfseries Atlas} Collaboration, G.~Aad {\em et~al.}, ``{Observation of a
  Centrality-Dependent Dijet Asymmetry in Lead-Lead Collisions at
  $\sqrt{s_{NN}}$= 2.76 TeV with the ATLAS Detector at the LHC},'' {\em
  Phys.Rev.Lett.} {\bfseries 105} (2010) 252303,
  \href{http://arxiv.org/abs/1011.6182}{{\ttfamily arXiv:1011.6182 [hep-ex]}}.

\bibitem{Chatrchyan:2012nia}
{\bfseries CMS} Collaboration, S.~Chatrchyan {\em et~al.}, ``{Jet momentum
  dependence of jet quenching in Pb--Pb collisions at $\sqrt{s_{NN}}=2.76$
  TeV},'' \href{http://dx.doi.org/10.1016/j.physletb.2012.04.058}{{\em
  Phys.Lett.} {\bfseries B712} (2012) 176--197},
\href{http://arxiv.org/abs/1202.5022}{{\ttfamily arXiv:1202.5022 [nucl-ex]}}.

\bibitem{Aad:2012vca}
{\bfseries ATLAS} Collaboration, G.~Aad {\em et~al.}, ``{Measurement of the jet
  radius and transverse momentum dependence of inclusive jet suppression in
  lead-lead collisions at $\sqrt{s_{NN}}=2.76$ TeV with the ATLAS detector},''
  \href{http://dx.doi.org/10.1016/j.physletb.2013.01.024}{{\em Phys.Lett.}
  {\bfseries B719} (2013) 220--241},
\href{http://arxiv.org/abs/1208.1967}{{\ttfamily arXiv:1208.1967 [hep-ex]}}.

\bibitem{Aamodt:2008zz}
{\bfseries ALICE} Collaboration, K.~Aamodt {\em et~al.}, ``{The ALICE
  experiment at the CERN LHC},''
{\em JINST} {\bfseries 0803} (2008) S08002.

\bibitem{Abelev:2013qoq}
{\bfseries ALICE} Collaboration, B.~Abelev {\em et~al.}, ``{Centrality
  determination of Pb-Pb collisions at sqrt(sNN) = 2.76 TeV with ALICE},''
  \href{http://dx.doi.org/10.1103/PhysRevC.88.044909}{{\em Phys.Rev.}
  {\bfseries C88} (2013) 044909},
\href{http://arxiv.org/abs/1301.4361}{{\ttfamily arXiv:1301.4361 [nucl-ex]}}.

\bibitem{Alme:2010ke}
J.~Alme, Y.~Andres, H.~Appelshauser, S.~Bablok, N.~Bialas, {\em et~al.}, ``{The
  ALICE TPC, a large 3-dimensional tracking device with fast readout for
  ultra-high multiplicity events},''
  \href{http://dx.doi.org/10.1016/j.nima.2010.04.042}{{\em Nucl.Instrum.Meth.}
  {\bfseries A622} (2010) 316--367},
\href{http://arxiv.org/abs/1001.1950}{{\ttfamily arXiv:1001.1950
  [physics.ins-det]}}.

\bibitem{Aamodt:2010aa}
{\bfseries ALICE} Collaboration, K.~Aamodt {\em et~al.}, ``{Alignment of the
  ALICE Inner Tracking System with cosmic-ray tracks},''
  \href{http://dx.doi.org/10.1088/1748-0221/5/03/P03003}{{\em JINST} {\bfseries
  5} (2010) P03003},
\href{http://arxiv.org/abs/1001.0502}{{\ttfamily arXiv:1001.0502
  [physics.ins-det]}}.

\bibitem{Abelev:2012eq}
{\bfseries ALICE} Collaboration, B.~Abelev {\em et~al.}, ``{Centrality
  Dependence of Charged Particle Production at Large Transverse Momentum in
  Pb--Pb Collisions at $\sqrt{s_{\rm{NN}}} = 2.76$ TeV},''
  \href{http://dx.doi.org/10.1016/j.physletb.2013.01.051}{{\em Phys.Lett.}
  {\bfseries B720} (2013) 52--62},
\href{http://arxiv.org/abs/1208.2711}{{\ttfamily arXiv:1208.2711 [hep-ex]}}.

\bibitem{Cacciari2011}
M.~Cacciari, G.~P. Salam, and G.~Soyez, ``{FastJet User Manual},''
  \href{http://dx.doi.org/10.1140/epjc/s10052-012-1896-2}{{\em Eur.Phys.J.}
  {\bfseries C72} (2012) 1896},
\href{http://arxiv.org/abs/1111.6097}{{\ttfamily arXiv:1111.6097 [hep-ph]}}.

\bibitem{Cacciari2006}
M.~Cacciari and G.~P. Salam, ``Dispelling the $n^3$ myth for the kt
  jet-finder,'' {\em Phys.Lett.B} {\bfseries 641} (2006) 57--61,
  \href{http://arxiv.org/abs/hep-ph/0512210}{{\ttfamily hep-ph/0512210}}.

\bibitem{Cacciari2008a}
M.~Cacciari, G.~P. Salam, and G.~Soyez, ``The catchment area of jets,'' {\em
  JHEP} {\bfseries 04} (Feb., 2008) 005,
  \href{http://arxiv.org/abs/0802.1188}{{\ttfamily 0802.1188}}.

\bibitem{Cacciari2008}
M.~Cacciari and G.~P. Salam, ``Pileup subtraction using jet areas,'' {\em
  Phys.Lett.B} {\bfseries 659} (2008) 119--126,
  \href{http://arxiv.org/abs/0707.1378}{{\ttfamily 0707.1378}}.

\bibitem{Cacciari2010}
M.~Cacciari, J.~Rojo, G.~P. Salam, and G.~Soyez, ``Jet reconstruction in heavy
  ion collisions,'' {\em Eur.Phys.J.C} {\bfseries 71} (Oct., 2010) 1539,
  \href{http://arxiv.org/abs/1010.1759}{{\ttfamily 1010.1759}}.

\bibitem{Abelev:2012ej}
{\bfseries ALICE} Collaboration, B.~Abelev {\em et~al.}, ``{Measurement of
  Event Background Fluctuations for Charged Particle Jet Reconstruction in
  Pb-Pb collisions at $\sqrt{s_{NN}} = 2.76$ TeV},''
  \href{http://dx.doi.org/10.1007/JHEP03(2012)053}{{\em JHEP} {\bfseries 1203}
  (2012) 053},
\href{http://arxiv.org/abs/1201.2423}{{\ttfamily arXiv:1201.2423 [hep-ex]}}.

\bibitem{Monk:2011pg}
J.~W. Monk and C.~Oropeza-Barrera, ``{The HBOM Method for Unfolding Detector
  Effects},'' \href{http://dx.doi.org/10.1016/j.nima.2012.09.045}{{\em
  Nucl.Instrum.Meth.} {\bfseries A701} (2013) 17--24},
\href{http://arxiv.org/abs/1111.4896}{{\ttfamily arXiv:1111.4896 [hep-ex]}}.

\bibitem{deBarros:2012ws}
G.~de~Barros, B.~Fenton-Olsen, P.~Jacobs, and M.~Ploskon, ``{Data-driven
  analysis methods for the measurement of reconstructed jets in heavyion
  collisions at RHIC and LHC },''
\href{http://arxiv.org/abs/1208.1518}{{\ttfamily arXiv:1208.1518 [hep-ex]}}.

\bibitem{Cacciari:2010te}
M.~Cacciari, J.~Rojo, G.~P. Salam, and G.~Soyez, ``{Jet Reconstruction in Heavy
  Ion Collisions},''
  \href{http://dx.doi.org/10.1140/epjc/s10052-011-1539-z}{{\em Eur.Phys.J.}
  {\bfseries C71} (2011) 1539},
\href{http://arxiv.org/abs/1010.1759}{{\ttfamily arXiv:1010.1759 [hep-ph]}}.

\bibitem{Sjostrand2006}
T.~Sjostrand, S.~Mrenna, and P.~Skands, ``{PYTHIA} 6.4 physics and manual,''
  {\em JHEP} {\bfseries 05} (2006) 026,
  \href{http://arxiv.org/abs/hep-ph/0603175}{{\ttfamily hep-ph/0603175}}.

\bibitem{Brun1994}
R.~Brun {\em et~al.}, {\em {GEANT} Detector Description and Simulation Tool},
  {CERN} program library long writeup w5013~ed., March, 1994.

\bibitem{Wang:1991hta}
X.-N. Wang and M.~Gyulassy, ``{HIJING: A Monte Carlo model for multiple jet
  production in p p, p A and A A collisions},''
\href{http://dx.doi.org/10.1103/PhysRevD.44.3501}{{\em Phys.Rev.} {\bfseries
  D44} (1991) 3501--3516}.

\bibitem{Skands:2010ak}
P.~Z. Skands, ``{Tuning Monte Carlo Generators: The Perugia Tunes},''
  \href{http://dx.doi.org/10.1103/PhysRevD.82.074018}{{\em Phys.Rev.}
  {\bfseries D82} (2010) 074018},
\href{http://arxiv.org/abs/1005.3457}{{\ttfamily arXiv:1005.3457 [hep-ph]}}.

\bibitem{tagkey2013262}
``Measurement of the inclusive differential jet cross section in pp collisions
  at,''
  \href{http://dx.doi.org/http://dx.doi.org/10.1016/j.physletb.2013.04.026}{{\em
  Physics Letters B} {\bfseries 722} no.~4-5, (2013) 262--272}.
  \url{http://www.sciencedirect.com/science/article/pii/S0370269313003055}.

\bibitem{Vajzer:2013gla}
{\bfseries ALICE} Collaboration, M.~Vajzer, ``{Charged jet spectra in
  proton-proton collisions with the ALICE experiment at the LHC},''
\href{http://arxiv.org/abs/1311.0148}{{\ttfamily arXiv:1311.0148 [hep-ex]}}.

\bibitem{Abelev:2012cn}
{\bfseries ALICE} Collaboration, B.~Abelev {\em et~al.}, ``{Neutral pion and
  $\eta$ meson production in proton-proton collisions at $\sqrt{s}=0.9$ TeV and
  $\sqrt{s}=7$ TeV},''
  \href{http://dx.doi.org/10.1016/j.physletb.2012.09.015}{{\em Phys.Lett.}
  {\bfseries B717} (2012) 162--172},
\href{http://arxiv.org/abs/1205.5724}{{\ttfamily arXiv:1205.5724 [hep-ex]}}.

\bibitem{D'Agostini:1994zf}
G.~D'Agostini, ``{A Multidimensional unfolding method based on Bayes'
  theorem},''
\href{http://dx.doi.org/10.1016/0168-9002(95)00274-X}{{\em Nucl.Instrum.Meth.}
  {\bfseries A362} (1995) 487--498}.

\bibitem{Blobel:2002pu}
V.~Blobel, ``{An Unfolding method for high-energy physics experiments },''
\href{http://arxiv.org/abs/hep-ex/0208022}{{\ttfamily arXiv:hep-ex/0208022
  [hep-ex]}}.

\bibitem{Schmelling1994400}
M.~Schmelling, ``The method of reduced cross-entropy a general approach to
  unfold probability distributions,''
  \href{http://dx.doi.org/10.1016/0168-9002(94)90119-8}{{\em Nuclear
  Instruments and Methods in Physics Research Section A: Accelerators,
  Spectrometers, Detectors and Associated Equipment} {\bfseries 340} no.~2,
  (1994) 400 -- 412}.
  \url{http://www.sciencedirect.com/science/article/pii/0168900294901198}.

\bibitem{Blobel1984}
V.~Blobel, {\em 8th CERN School of Computing - CSC '84, Aiguablava, Spain 9--22
  Sep.}, 1984.

\bibitem{dagostini:2003}
G.~D'Agostini, ``{Bayesian inference in processing experimental data:
  Principles and basic applications},''
  \href{http://dx.doi.org/10.1088/0034-4885/66/9/201}{{\em Rept.Prog.Phys.}
  {\bfseries 66} (2003) 1383--1420},
\href{http://arxiv.org/abs/physics/0304102}{{\ttfamily arXiv:physics/0304102
  [physics]}}.

\bibitem{Hocker:1995kb}
A.~Hocker and V.~Kartvelishvili, ``{SVD approach to data unfolding},''
  \href{http://dx.doi.org/10.1016/0168-9002(95)01478-0}{{\em
  Nucl.Instrum.Meth.} {\bfseries A372} (1996) 469--481},
\href{http://arxiv.org/abs/hep-ph/9509307}{{\ttfamily arXiv:hep-ph/9509307
  [hep-ph]}}.

\bibitem{Adye:2011gm}
T.~Adye, ``{Unfolding algorithms and tests using RooUnfold },''
\href{http://arxiv.org/abs/1105.1160}{{\ttfamily arXiv:1105.1160
  [physics.data-an]}}.

\bibitem{Amaudruz:1995tq}
{\bfseries New Muon} Collaboration, P.~Amaudruz {\em et~al.}, ``{A Reevaluation
  of the nuclear structure function ratios for D, He, Li-6, C and Ca},''
  \href{http://dx.doi.org/10.1016/0550-3213(94)00023-9}{{\em Nucl.Phys.}
  {\bfseries B441} (1995) 3--11},
\href{http://arxiv.org/abs/hep-ph/9503291}{{\ttfamily arXiv:hep-ph/9503291
  [hep-ph]}}.

\bibitem{Eskola:2009uj}
K.~Eskola, H.~Paukkunen, and C.~Salgado, ``{EPS09: A New Generation of NLO and
  LO Nuclear Parton Distribution Functions},''
  \href{http://dx.doi.org/10.1088/1126-6708/2009/04/065}{{\em JHEP} {\bfseries
  0904} (2009) 065},
\href{http://arxiv.org/abs/0902.4154}{{\ttfamily arXiv:0902.4154 [hep-ph]}}.

\bibitem{Cronin:1974zm}
J.~Cronin, H.~J. Frisch, M.~Shochet, J.~Boymond, R.~Mermod, {\em et~al.},
  ``{Production of Hadrons with Large Transverse Momentum at 200-GeV, 300-GeV,
  and 400-GeV},''
\href{http://dx.doi.org/10.1103/PhysRevD.11.3105}{{\em Phys.Rev.} {\bfseries
  D11} (1975) 3105}.

\bibitem{Vitev:2008bx}
I.~Vitev, B.-W. Zhang, and S.~Wicks, ``{The Theory and phenomenology of jets in
  nuclear collisions},''
  \href{http://dx.doi.org/10.1140/epjc/s10052-009-1025-z}{{\em Eur.Phys.J.}
  {\bfseries C62} (2009) 139--144},
\href{http://arxiv.org/abs/0810.3052}{{\ttfamily arXiv:0810.3052 [hep-ph]}}.

\bibitem{Vitev:2008rz}
I.~Vitev, S.~Wicks, and B.-W. Zhang, ``{A Theory of jet shapes and cross
  sections: From hadrons to nuclei},''
  \href{http://dx.doi.org/10.1088/1126-6708/2008/11/093}{{\em JHEP} {\bfseries
  0811} (2008) 093},
\href{http://arxiv.org/abs/0810.2807}{{\ttfamily arXiv:0810.2807 [hep-ph]}}.

\bibitem{Salgado:2003gb}
C.~A. Salgado and U.~A. Wiedemann, ``{Calculating quenching weights},''
  \href{http://dx.doi.org/10.1103/PhysRevD.68.014008}{{\em Phys. Rev.}
  {\bfseries D68} (2003) 014008},
\href{http://arxiv.org/abs/hep-ph/0302184}{{\ttfamily arXiv:hep-ph/0302184}}.

\bibitem{Vitev:2005yg}
I.~Vitev, ``{Large angle hadron correlations from medium-induced gluon
  radiation},'' \href{http://dx.doi.org/10.1016/j.physletb.2005.09.082}{{\em
  Phys.Lett.} {\bfseries B630} (2005) 78--84},
\href{http://arxiv.org/abs/hep-ph/0501255}{{\ttfamily arXiv:hep-ph/0501255
  [hep-ph]}}.

\bibitem{Armesto:2009fj}
N.~Armesto, L.~Cunqueiro, and C.~A. Salgado, ``{Q-PYTHIA: A Medium-modified
  implementation of final state radiation},''
  \href{http://dx.doi.org/10.1140/epjc/s10052-009-1133-9}{{\em Eur.Phys.J.}
  {\bfseries C63} (2009) 679--690},
\href{http://arxiv.org/abs/0907.1014}{{\ttfamily arXiv:0907.1014 [hep-ph]}}.

\bibitem{Apolinario:2012cg}
L.~Apolinario, N.~Armesto, and L.~Cunqueiro, ``{An analysis of the influence of
  background subtraction and quenching on jet observables in heavy-ion
  collisions},'' \href{http://dx.doi.org/10.1007/JHEP02(2013)022}{{\em JHEP}
  {\bfseries 1302} (2013) 022},
\href{http://arxiv.org/abs/1211.1161}{{\ttfamily arXiv:1211.1161 [hep-ph]}}.

\bibitem{Chatrchyan:2011sx}
{\bfseries CMS} Collaboration, S.~Chatrchyan {\em et~al.}, ``{Observation and
  studies of jet quenching in PbPb collisions at nucleon-nucleon center-of-mass
  energy = 2.76 TeV},''
  \href{http://dx.doi.org/10.1103/PhysRevC.84.024906}{{\em Phys.Rev.}
  {\bfseries C84} (2011) 024906},
\href{http://arxiv.org/abs/1102.1957}{{\ttfamily arXiv:1102.1957 [nucl-ex]}}.

\bibitem{Renk:2009nz}
T.~Renk, ``{A Comparison study of medium-modified QCD shower evolution
  scenarios},'' \href{http://dx.doi.org/10.1103/PhysRevC.79.054906}{{\em
  Phys.Rev.} {\bfseries C79} (2009) 054906},
\href{http://arxiv.org/abs/0901.2818}{{\ttfamily arXiv:0901.2818 [hep-ph]}}.

\bibitem{Renk:2010zx}
T.~Renk, ``{YaJEM: a Monte Carlo code for in-medium shower evolution},''
  \href{http://dx.doi.org/10.1142/S0218301311019933}{{\em Int.J.Mod.Phys.}
  {\bfseries E20} (2011) 1594--1599},
\href{http://arxiv.org/abs/1009.3740}{{\ttfamily arXiv:1009.3740 [hep-ph]}}.

\bibitem{CasalderreySolana:2010eh}
J.~Casalderrey-Solana, J.~G. Milhano, and U.~A. Wiedemann, ``{Jet Quenching via
  Jet Collimation},''
  \href{http://dx.doi.org/10.1088/0954-3899/38/3/035006}{{\em J.Phys.}
  {\bfseries G38} (2011) 035006},
\href{http://arxiv.org/abs/1012.0745}{{\ttfamily arXiv:1012.0745 [hep-ph]}}.

\bibitem{Renk:2012cx}
T.~Renk, ``{On the sensitivity of the dijet asymmetry to the physics of jet
  quenching},'' \href{http://dx.doi.org/10.1103/PhysRevC.85.064908}{{\em
  Phys.Rev.} {\bfseries C85} (2012) 064908},
\href{http://arxiv.org/abs/1202.4579}{{\ttfamily arXiv:1202.4579 [hep-ph]}}.

\bibitem{MehtarTani:2012cy}
Y.~Mehtar-Tani, C.~A. Salgado, and K.~Tywoniuk, ``{The Radiation pattern of a
  QCD antenna in a dense medium},''
  \href{http://dx.doi.org/10.1007/JHEP10(2012)197}{{\em JHEP} {\bfseries 1210}
  (2012) 197},
\href{http://arxiv.org/abs/1205.5739}{{\ttfamily arXiv:1205.5739 [hep-ph]}}.

\bibitem{MehtarTani:2011tz}
Y.~Mehtar-Tani, C.~Salgado, and K.~Tywoniuk, ``{Jets in QCD Media: From Color
  Coherence to Decoherence},''
  \href{http://dx.doi.org/10.1016/j.physletb.2011.12.042}{{\em Phys.Lett.}
  {\bfseries B707} (2012) 156--159},
\href{http://arxiv.org/abs/1102.4317}{{\ttfamily arXiv:1102.4317 [hep-ph]}}.

\bibitem{Chatrchyan:2012gw}
{\bfseries CMS} Collaboration, S.~Chatrchyan {\em et~al.}, ``{Measurement of
  jet fragmentation into charged particles in $pp$ and PbPb collisions at
  $\sqrt{s_{NN}}=2.76$ TeV},''
  \href{http://dx.doi.org/10.1007/JHEP10(2012)087}{{\em JHEP} {\bfseries 1210}
  (2012) 087},
\href{http://arxiv.org/abs/1205.5872}{{\ttfamily arXiv:1205.5872 [nucl-ex]}}.

\bibitem{Chatrchyan:2013kwa}
{\bfseries CMS} Collaboration, S.~Chatrchyan {\em et~al.}, ``{Modification of
  jet shapes in PbPb collisions at $\sqrt{s_{NN}}$ = 2.76 TeV},''
\href{http://arxiv.org/abs/1310.0878}{{\ttfamily arXiv:1310.0878 [nucl-ex]}}.

\bibitem{ATLAS_ff_conf}
{\bfseries ATLAS collaboration} Collaboration, G.~Aad {\em et~al.}

\bibitem{CasalderreySolana:2011gx}
J.~Casalderrey-Solana, J.~G. Milhano, and P.~Q. Arias, ``{Out of Medium
  Fragmentation from Long-Lived Jet Showers},''
  \href{http://dx.doi.org/10.1016/j.physletb.2012.02.066}{{\em Phys.Lett.}
  {\bfseries B710} (2012) 175--181},
\href{http://arxiv.org/abs/1111.0310}{{\ttfamily arXiv:1111.0310 [hep-ph]}}.

\bibitem{Mehtar-Tani:2013pia}
Y.~Mehtar-Tani, J.~Milhano, and K.~Tywoniuk, ``{Jet physics in heavy-ion
  collisions},'' \href{http://dx.doi.org/10.1142/S0217751X13400137}{{\em
  Int.J.Mod.Phys.} {\bfseries A28} (2013) 1340013},
\href{http://arxiv.org/abs/1302.2579}{{\ttfamily arXiv:1302.2579 [hep-ph]}}.

\bibitem{Renk:2012ve}
T.~Renk, ``{Biased Showers - a common conceptual Framework for the
  Interpretation of High pT Observables in Heavy-Ion Collisions },''
\href{http://arxiv.org/abs/1212.0646}{{\ttfamily arXiv:1212.0646 [hep-ph]}}.

\bibitem{Zapp:2012ak}
K.~C. Zapp, F.~Krauss, and U.~A. Wiedemann, ``{A perturbative framework for jet
  quenching},'' \href{http://dx.doi.org/10.1007/JHEP03(2013)080}{{\em JHEP}
  {\bfseries 1303} (2013) 080},
\href{http://arxiv.org/abs/1212.1599}{{\ttfamily arXiv:1212.1599 [hep-ph]}}.

\end{thebibliography}\endgroup

\appendix

\section{\chisq{} minimization unfolding method}\label{app:UnfMethods}
The \chisq{} minimization method minimizes the difference between the refolded and measured spectrum \cite{Blobel1984}. The refolded spectrum is the unfolded distribution convoluted with the response matrix. The $\chi^{2}$ function to be minimized indicates how well the refolded distribution describes the measured spectrum:
\begin{equation}\label{eq:chi2functionGen}
\chi^{2}_{\rom{fit}} =
\displaystyle\sum_{\mathrm{refolded}}\left(\frac{y_{\mathrm{refolded}}-y_{\mathrm{measured}}}{\sigma_{\mathrm{measured}}}\right)^{2},
\end{equation}
in which $y$ is the yield of the refolded or measured distribution and $\sigma_{\rom{measured}}$ the statistical uncertainty on the measured distribution. 
The true distribution minimizes this \chisq{} function but in addition also many other fluctuating solutions exist. Heavily fluctuating solutions can be damped by adding a penalty term to the \chisq{} function:
\begin{equation}\label{eq:chi2functionGenPen}
\chi^{2} =
\displaystyle\sum_{\mathrm{refolded}}\left(\frac{y_{\mathrm{refolded}}-y_{\mathrm{measured}}}{\sigma_{\mathrm{measured}}}\right)^{2}
+ \beta P(y_{\mathrm{unfolded}}),
\end{equation}
where $y_{\mathrm{unfolded}}$ is the unfolded distribution.
$\beta P(y_{\mathrm{unfolded}})$ is the penalty term which regularizes the unfolded distribution. The strength of the applied regularization is given by $\beta$ and $P(y_{\mathrm{unfolded}})$ is the regularization term favoring a certain shape. The choice of the regularization function is motivated by the expected shape of the solution. For this analysis the regularization favors a local power law which is calculated using finite differences:
\begin{equation}
 P(y_{\mathrm{unfolded}}) = \displaystyle\sum_{\mathrm{unfolded}}\left(\frac{\mathrm{d}^{2}\log y_{\mathrm{unfolded}}}{\mathrm{d}
   \log \pt^{2}}\right) ^{2}.
\end{equation}
Note that the exponent in the power law is not fixed and is not required to be the same over the full unfolded solution.
Sensitivity of the unfolded distribution to this particular choice of regularization can be tested by varying the regularization strength $\beta$ and by comparing the unfolded distribution to a solution with a different functional shape for the regularization.

In case the regularization is dominant the penalty term is of the same order or larger than the $\chi^{2}_{\rom{fit}}$ between the refolded and measured spectrum. In this case the refolded spectrum does not describe the measured spectrum and the $\chi^{2}_{\rom{fit}}$ between the refolded and measured spectrum is large.

The covariance matrix for the unfolded spectrum is calculated in the usual way, by inverting the Hessian matrix.
In case the regularization is too weak or too strong, off-diagonal correlations in the Pearson coefficients extracted from the covariance matrix appear.

\newpage

\section{The ALICE Collaboration}\label{app:collab}



\begingroup
\small
\begin{flushleft}
B.~Abelev\Irefn{org74}\And
J.~Adam\Irefn{org38}\And
D.~Adamov\'{a}\Irefn{org82}\And
M.M.~Aggarwal\Irefn{org86}\And
G.~Aglieri~Rinella\Irefn{org34}\And
M.~Agnello\Irefn{org92}\textsuperscript{,}\Irefn{org109}\And
A.G.~Agocs\Irefn{org132}\And
A.~Agostinelli\Irefn{org26}\And
N.~Agrawal\Irefn{org45}\And
Z.~Ahammed\Irefn{org128}
% ...
\renewcommand\labelenumi{\textsuperscript{\theenumi}~}

\section*{Affiliation notes}
\renewcommand\theenumi{\roman{enumi}}
\begin{Authlist}
\item \Adef{0}Deceased
\item \Adef{idp140722642658848}{Also at: St-Petersburg State Polytechnical University}
\item \Adef{idp140722644517424}{Also at: Department of Applied Physics, Aligarh Muslim University, Aligarh, India}
\item \Adef{idp140722654652448}{Also at: M.V. Lomonosov Moscow State University, D.V. Skobeltsyn Institute of Nuclear Physics, Moscow, Russia}
\item \Adef{idp140722654895248}{Also at: University of Belgrade, Faculty of Physics and "Vin\v{c}a" Institute of Nuclear Sciences, Belgrade, Serbia}
\item \Adef{idp140722655174576}{Permanent address: Konkuk University, Seoul, Korea}
\item \Adef{idp140722655704576}{Also at: Institute of Theoretical Physics, University of Wroclaw, Wroclaw, Poland}
\item \Adef{idp140722656596992}{Also at: the University of Kansas, Lawrence, KS, United States}
\end{Authlist}

\section*{Collaboration Institutes}
\renewcommand\theenumi{\arabic{enumi}~}
\begin{Authlist}

\item \Idef{org1}A.I. Alikhanyan National Science Laboratory (Yerevan Physics Institute) Foundation, Yerevan, Armenia
\item \Idef{org2}Benem\'{e}rita Universidad Aut\'{o}noma de Puebla, Puebla, Mexico
\item \Idef{org3}Bogolyubov Institute for Theoretical Physics, Kiev, Ukraine
\item \Idef{org4}Bose Institute, Department of Physics and Centre for Astroparticle Physics and Space Science (CAPSS), Kolkata, India
\item \Idef{org5}Budker Institute for Nuclear Physics, Novosibirsk, Russia
\item \Idef{org6}California Polytechnic State University, San Luis Obispo, CA, United States
\item \Idef{org7}Central China Normal University, Wuhan, China
\item \Idef{org8}Centre de Calcul de l'IN2P3, Villeurbanne, France
\item \Idef{org9}Centro de Aplicaciones Tecnol\'{o}gicas y Desarrollo Nuclear (CEADEN), Havana, Cuba
\item \Idef{org10}Centro de Investigaciones Energ\'{e}ticas Medioambientales y Tecnol\'{o}gicas (CIEMAT), Madrid, Spain
\item \Idef{org11}Centro de Investigaci\'{o}n y de Estudios Avanzados (CINVESTAV), Mexico City and M\'{e}rida, Mexico
\item \Idef{org12}Centro Fermi - Museo Storico della Fisica e Centro Studi e Ricerche ``Enrico Fermi'', Rome, Italy
\item \Idef{org13}Chicago State University, Chicago, USA
\item \Idef{org14}Commissariat \`{a} l'Energie Atomique, IRFU, Saclay, France
\item \Idef{org15}COMSATS Institute of Information Technology (CIIT), Islamabad, Pakistan
\item \Idef{org16}Departamento de F\'{\i}sica de Part\'{\i}culas and IGFAE, Universidad de Santiago de Compostela, Santiago de Compostela, Spain
\item \Idef{org17}Department of Physics and Technology, University of Bergen, Bergen, Norway
\item \Idef{org18}Department of Physics, Aligarh Muslim University, Aligarh, India
\item \Idef{org19}Department of Physics, Ohio State University, Columbus, OH, United States
\item \Idef{org20}Department of Physics, Sejong University, Seoul, South Korea
\item \Idef{org21}Department of Physics, University of Oslo, Oslo, Norway
\item \Idef{org22}Dipartimento di Fisica dell'Universit\`{a} 'La Sapienza' and Sezione INFN Rome
\item \Idef{org23}Dipartimento di Fisica dell'Universit\`{a} and Sezione INFN, Cagliari, Italy
\item \Idef{org24}Dipartimento di Fisica dell'Universit\`{a} and Sezione INFN, Trieste, Italy
\item \Idef{org25}Dipartimento di Fisica dell'Universit\`{a} and Sezione INFN, Turin, Italy
\item \Idef{org26}Dipartimento di Fisica e Astronomia dell'Universit\`{a} and Sezione INFN, Bologna, Italy
\item \Idef{org27}Dipartimento di Fisica e Astronomia dell'Universit\`{a} and Sezione INFN, Catania, Italy
\item \Idef{org28}Dipartimento di Fisica e Astronomia dell'Universit\`{a} and Sezione INFN, Padova, Italy
\item \Idef{org29}Dipartimento di Fisica `E.R.~Caianiello' dell'Universit\`{a} and Gruppo Collegato INFN, Salerno, Italy
\item \Idef{org30}Dipartimento di Scienze e Innovazione Tecnologica dell'Universit\`{a} del  Piemonte Orientale and Gruppo Collegato INFN, Alessandria, Italy
\item \Idef{org31}Dipartimento Interateneo di Fisica `M.~Merlin' and Sezione INFN, Bari, Italy
\item \Idef{org32}Division of Experimental High Energy Physics, University of Lund, Lund, Sweden
\item \Idef{org33}Eberhard Karls Universit\"{a}t T\"{u}bingen, T\"{u}bingen, Germany
\item \Idef{org34}European Organization for Nuclear Research (CERN), Geneva, Switzerland
\item \Idef{org35}Excellence Cluster Universe, Technische Universit\"{a}t M\"{u}nchen, Munich, Germany
\item \Idef{org36}Faculty of Engineering, Bergen University College, Bergen, Norway
\item \Idef{org37}Faculty of Mathematics, Physics and Informatics, Comenius University, Bratislava, Slovakia
\item \Idef{org38}Faculty of Nuclear Sciences and Physical Engineering, Czech Technical University in Prague, Prague, Czech Republic
\item \Idef{org39}Faculty of Science, P.J.~\v{S}af\'{a}rik University, Ko\v{s}ice, Slovakia
\item \Idef{org40}Frankfurt Institute for Advanced Studies, Johann Wolfgang Goethe-Universit\"{a}t Frankfurt, Frankfurt, Germany
\item \Idef{org41}Gangneung-Wonju National University, Gangneung, South Korea
\item \Idef{org42}Gauhati University, Department of Physics, Guwahati, India
\item \Idef{org43}Helsinki Institute of Physics (HIP), Helsinki, Finland
\item \Idef{org44}Hiroshima University, Hiroshima, Japan
\item \Idef{org45}Indian Institute of Technology Bombay (IIT), Mumbai, India
\item \Idef{org46}Indian Institute of Technology Indore, Indore (IITI), India
\item \Idef{org47}Inha University, College of Natural Sciences
\item \Idef{org48}Institut de Physique Nucleaire d'Orsay (IPNO), Universite Paris-Sud, CNRS-IN2P3, Orsay, France
\item \Idef{org49}Institut f\"{u}r Informatik, Johann Wolfgang Goethe-Universit\"{a}t Frankfurt, Frankfurt, Germany
\item \Idef{org50}Institut f\"{u}r Kernphysik, Johann Wolfgang Goethe-Universit\"{a}t Frankfurt, Frankfurt, Germany
\item \Idef{org51}Institut f\"{u}r Kernphysik, Technische Universit\"{a}t Darmstadt, Darmstadt, Germany
\item \Idef{org52}Institut f\"{u}r Kernphysik, Westf\"{a}lische Wilhelms-Universit\"{a}t M\"{u}nster, M\"{u}nster, Germany
\item \Idef{org53}Institut Pluridisciplinaire Hubert Curien (IPHC), Universit\'{e} de Strasbourg, CNRS-IN2P3, Strasbourg, France
\item \Idef{org54}Institute for High Energy Physics, Protvino, Russia
\item \Idef{org55}Institute for Nuclear Research, Academy of Sciences, Moscow, Russia
\item \Idef{org56}Institute for Subatomic Physics of Utrecht University, Utrecht, Netherlands
\item \Idef{org57}Institute for Theoretical and Experimental Physics, Moscow, Russia
\item \Idef{org58}Institute of Experimental Physics, Slovak Academy of Sciences, Ko\v{s}ice, Slovakia
\item \Idef{org59}Institute of Physics, Academy of Sciences of the Czech Republic, Prague, Czech Republic
\item \Idef{org60}Institute of Physics, Bhubaneswar, India
\item \Idef{org61}Institute of Space Science (ISS), Bucharest, Romania
\item \Idef{org62}Instituto de Ciencias Nucleares, Universidad Nacional Aut\'{o}noma de M\'{e}xico, Mexico City, Mexico
\item \Idef{org63}Instituto de F\'{\i}sica, Universidad Nacional Aut\'{o}noma de M\'{e}xico, Mexico City, Mexico
\item \Idef{org64}iThemba LABS, National Research Foundation, Somerset West, South Africa
\item \Idef{org65}Joint Institute for Nuclear Research (JINR), Dubna, Russia
\item \Idef{org66}Konkuk University
\item \Idef{org67}Korea Institute of Science and Technology Information, Daejeon, South Korea
\item \Idef{org68}KTO Karatay University, Konya, Turkey
\item \Idef{org69}Laboratoire de Physique Corpusculaire (LPC), Clermont Universit\'{e}, Universit\'{e} Blaise Pascal, CNRS--IN2P3, Clermont-Ferrand, France
\item \Idef{org70}Laboratoire de Physique Subatomique et de Cosmologie (LPSC), Universit\'{e} Joseph Fourier, CNRS-IN2P3, Institut Polytechnique de Grenoble, Grenoble, France
\item \Idef{org71}Laboratori Nazionali di Frascati, INFN, Frascati, Italy
\item \Idef{org72}Laboratori Nazionali di Legnaro, INFN, Legnaro, Italy
\item \Idef{org73}Lawrence Berkeley National Laboratory, Berkeley, CA, United States
\item \Idef{org74}Lawrence Livermore National Laboratory, Livermore, CA, United States
\item \Idef{org75}Moscow Engineering Physics Institute, Moscow, Russia
\item \Idef{org76}National Centre for Nuclear Studies, Warsaw, Poland
\item \Idef{org77}National Institute for Physics and Nuclear Engineering, Bucharest, Romania
\item \Idef{org78}National Institute of Science Education and Research, Bhubaneswar, India
\item \Idef{org79}Niels Bohr Institute, University of Copenhagen, Copenhagen, Denmark
\item \Idef{org80}Nikhef, National Institute for Subatomic Physics, Amsterdam, Netherlands
\item \Idef{org81}Nuclear Physics Group, STFC Daresbury Laboratory, Daresbury, United Kingdom
\item \Idef{org82}Nuclear Physics Institute, Academy of Sciences of the Czech Republic, \v{R}e\v{z} u Prahy, Czech Republic
\item \Idef{org83}Oak Ridge National Laboratory, Oak Ridge, TN, United States
\item \Idef{org84}Petersburg Nuclear Physics Institute, Gatchina, Russia
\item \Idef{org85}Physics Department, Creighton University, Omaha, NE, United States
\item \Idef{org86}Physics Department, Panjab University, Chandigarh, India
\item \Idef{org87}Physics Department, University of Athens, Athens, Greece
\item \Idef{org88}Physics Department, University of Cape Town, Cape Town, South Africa
\item \Idef{org89}Physics Department, University of Jammu, Jammu, India
\item \Idef{org90}Physics Department, University of Rajasthan, Jaipur, India
\item \Idef{org91}Physikalisches Institut, Ruprecht-Karls-Universit\"{a}t Heidelberg, Heidelberg, Germany
\item \Idef{org92}Politecnico di Torino, Turin, Italy
\item \Idef{org93}Purdue University, West Lafayette, IN, United States
\item \Idef{org94}Pusan National University, Pusan, South Korea
\item \Idef{org95}Research Division and ExtreMe Matter Institute EMMI, GSI Helmholtzzentrum f\"ur Schwerionenforschung, Darmstadt, Germany
\item \Idef{org96}Rudjer Bo\v{s}kovi\'{c} Institute, Zagreb, Croatia
\item \Idef{org97}Russian Federal Nuclear Center (VNIIEF), Sarov, Russia
\item \Idef{org98}Russian Research Centre Kurchatov Institute, Moscow, Russia
\item \Idef{org99}Saha Institute of Nuclear Physics, Kolkata, India
\item \Idef{org100}School of Physics and Astronomy, University of Birmingham, Birmingham, United Kingdom
\item \Idef{org101}Secci\'{o}n F\'{\i}sica, Departamento de Ciencias, Pontificia Universidad Cat\'{o}lica del Per\'{u}, Lima, Peru
\item \Idef{org102}Sezione INFN, Bari, Italy
\item \Idef{org103}Sezione INFN, Bologna, Italy
\item \Idef{org104}Sezione INFN, Cagliari, Italy
\item \Idef{org105}Sezione INFN, Catania, Italy
\item \Idef{org106}Sezione INFN, Padova, Italy
\item \Idef{org107}Sezione INFN, Rome, Italy
\item \Idef{org108}Sezione INFN, Trieste, Italy
\item \Idef{org109}Sezione INFN, Turin, Italy
\item \Idef{org110}SUBATECH, Ecole des Mines de Nantes, Universit\'{e} de Nantes, CNRS-IN2P3, Nantes, France
\item \Idef{org111}Suranaree University of Technology, Nakhon Ratchasima, Thailand
\item \Idef{org112}Technical University of Split FESB, Split, Croatia
\item \Idef{org113}Technische Universit\"{a}t M\"{u}nchen, Munich, Germany
\item \Idef{org114}The Henryk Niewodniczanski Institute of Nuclear Physics, Polish Academy of Sciences, Cracow, Poland
\item \Idef{org115}The University of Texas at Austin, Physics Department, Austin, TX, USA
\item \Idef{org116}Universidad Aut\'{o}noma de Sinaloa, Culiac\'{a}n, Mexico
\item \Idef{org117}Universidade de S\~{a}o Paulo (USP), S\~{a}o Paulo, Brazil
\item \Idef{org118}Universidade Estadual de Campinas (UNICAMP), Campinas, Brazil
\item \Idef{org119}University of Houston, Houston, TX, United States
\item \Idef{org120}University of Jyv\"{a}skyl\"{a}, Jyv\"{a}skyl\"{a}, Finland
\item \Idef{org121}University of Liverpool, Liverpool, United Kingdom
\item \Idef{org122}University of Tennessee, Knoxville, TN, United States
\item \Idef{org123}University of Tokyo, Tokyo, Japan
\item \Idef{org124}University of Tsukuba, Tsukuba, Japan
\item \Idef{org125}University of Zagreb, Zagreb, Croatia
\item \Idef{org126}Universit\'{e} de Lyon, Universit\'{e} Lyon 1, CNRS/IN2P3, IPN-Lyon, Villeurbanne, France
\item \Idef{org127}V.~Fock Institute for Physics, St. Petersburg State University, St. Petersburg, Russia
\item \Idef{org128}Variable Energy Cyclotron Centre, Kolkata, India
\item \Idef{org129}Vestfold University College, Tonsberg, Norway
\item \Idef{org130}Warsaw University of Technology, Warsaw, Poland
\item \Idef{org131}Wayne State University, Detroit, MI, United States
\item \Idef{org132}Wigner Research Centre for Physics, Hungarian Academy of Sciences, Budapest, Hungary
\item \Idef{org133}Yale University, New Haven, CT, United States
\item \Idef{org134}Yonsei University, Seoul, South Korea
\item \Idef{org135}Zentrum f\"{u}r Technologietransfer und Telekommunikation (ZTT), Fachhochschule Worms, Worms, Germany
\end{Authlist}
\endgroup


%
\end{document}